\documentclass[aps,pre,superscriptaddress,10pt,nofootinbib]{revtex4-1}

\usepackage[utf8]{inputenc}
\usepackage[english]{babel}

\usepackage{color}
\usepackage{graphicx}
\usepackage{amsmath}
\usepackage{amssymb}
\usepackage{mathrsfs}
\usepackage{fullpage}
\usepackage{relsize}
\usepackage{dcolumn}
\usepackage{longtable}

\newcommand{\alphaBare}{\alpha_0}
\newcommand{\dBare}{d_0}

\begin{document}

\title{Active-passive mixtures with bulk loading: a minimal active engine in one dimension}
\author{Jean-François Derivaux}
\affiliation{Department of Physics and Astronomy, University of Padova, Via Marzolo 8, 35131 Padova, Italy}
\affiliation{DAMTP, Centre for Mathematical Sciences, University of Cambridge,
Wilberforce Road, Cambridge CB3 0WA, United Kingdom}
\author{Robert L. Jack}
\affiliation{DAMTP, Centre for Mathematical Sciences, University of Cambridge,
Wilberforce Road, Cambridge CB3 0WA, United Kingdom}
\affiliation{Yusuf Hamied Department of Chemistry, University of Cambridge, Lensfield
Road, Cambridge CB2 1EW, United Kingdom}
\author{Michael E. Cates}
\affiliation{DAMTP, Centre for Mathematical Sciences, University of Cambridge,
Wilberforce Road, Cambridge CB3 0WA, United Kingdom}

\begin{abstract} 
We study a one-dimensional mixture of active (run-and-tumble) particles and passive (Brownian) particles, with single-file constraint, in a sawtooth  potential.  The active particles experience a ratchet effect: this generates a current, which can push passive particles against an applied load.  The resulting system operates as an active engine.  Using numerical simulations, we analyse the efficiency of this engine, and we discuss how it can be optimised.  Efficient operation occurs when the active particles self-organise into teams, which can push the passive ones against large loads by leveraging collective behaviour.  We discuss how the particle arrangement, conserved under the single-file constraint, affects the engine efficiency.  We also show that relaxing this constraint still allows the engine to operate effectively. 
\end{abstract}

\maketitle

\section{Introduction}

Active matter encompasses a diverse range of systems whose unifying feature is that their individual constituents are out of equilibrium, dissipating energy on a local scale \cite{ramaswamy_mechanics_2010,marchetti_hydrodynamics_2013}.
A large body of experimental and theoretical works have investigated the distinctive behaviour of such systems, which include bird flocks \cite{bialek_statistical_2012}, solutions of bacteria \cite{zhang_collective_2010,sokolov_physical_2012,elgeti_physics_2015}, the cell cytoskeleton \cite{prost_active_2015,julicher_hydrodynamic_2018} and catalytic self-propelling colloidal particles \cite{wang_small_2013,elgeti_physics_2015}. Compared with equilibrium systems, the collective behaviour of active matter is enriched by new phenomenology such as motility-induced phase separation \cite{cates_motility-induced_2015}, collective motion \cite{vicsek_collective_2012,bricard_emergence_2013}, active turbulence \cite{wensink_meso-scale_2012} and chemotaxis \cite{liebchen_synthetic_2018}.
Mixtures of active and passive objects support yet more phenomena, including enhanced diffusion of passive particles~\cite{leptos_dynamics_2009,zhao_enhanced_2017}, and novel instabilities of interfaces~\cite{wysocki_propagating_2016,stenhammar_activity-induced_2015,patteson_propagation_2018}  and bulk phases~\cite{wittkowski_nonequilibrium_2017,you_nonreciprocity_2020,saha_scalar_2020}.  
Active particles moving in {\em a-priori} random directions can also be rectified, by the presence of asymmetric obstacles or potentials, to create a steady current -- which is not possible in the passive case \cite{reichhardt_ratchet_2017}.

In the light of these results, active particles have been considered as a means of control for passive ones, for example in self-assembly of clusters~\cite{singh_non-equilibrium_2017,mallory_active_2018,niu_modular_2018,lowen_active_2018} or annealing of crystals~\cite{meer_fabricating_2016}. Mixtures of active and passive particles can also drive currents of the passive components \cite{ghosh_self-propelled_2013,bechinger_active_2016-1,reichhardt_ratchet_2017}. Such rectification has been illustrated in experiments \cite{koumakis_targeted_2013}, in which bacteria generated current of colloids through a sawtooth landscape. Transport of passive particles in an active/passive mixture can also be engineered by using non-mechanical forces, such as thermal gradients \cite{zhu_rectification_2020}.

Given such rectification effects, a natural extension is to develop active engines \cite{fodor_active_2021}, which harness the power generated by active constituents, in order to extract work.  One route towards such engines starts from (passive) Brownian heat engines \cite{imparato_work_2007,blickle_realization_2012,martinez_brownian_2016,martinez_colloidal_2016}, where colloidal beads are trapped by optical tweezers and subjected to cyclic change of external variables, such as the stiffness of the confining potential, or the temperature of the heat bath.  By adding active particles (such as bacteria) to the medium~\cite{martinez_colloidal_2016,krishnamurthy_micrometre-sized_2016,martinez_colloidal_2016}, the forces acting on the colloids become non-Gaussian and these engines become active~\cite{krishnamurthy_micrometre-sized_2016,zakine_stochastic_2017,holubec_active_2020}.  Alternatively, one may construct active engines based on the interactions of active particles with a wall~\cite{ekeh_thermodynamic_2020}.  These active engines can extract work from active particles under isothermal conditions.

A second type of active engine involves ratchet mechanisms: these generate active currents in the system, which are exploited to lift weights or push obstacles.  These engines are usually called \textit{autonomous}, in contrast with the engines requiring cyclic interventions as discussed above.  They can be realised by introducing asymmetric obstacles on which the active particles do work: the shape of the obstacles can be designed to optimise efficiency~\cite{pietzonka_autonomous_2019}.  Such engines have been realised experimentally, with bacteria generating the rotary motion of an asymmetric gearwheel \cite{sokolov_swimming_2010,leonardo_bacterial_2010,vizsnyiczai_light_2017}. 

\begin{figure}
\includegraphics[width = 0.75 \textwidth]{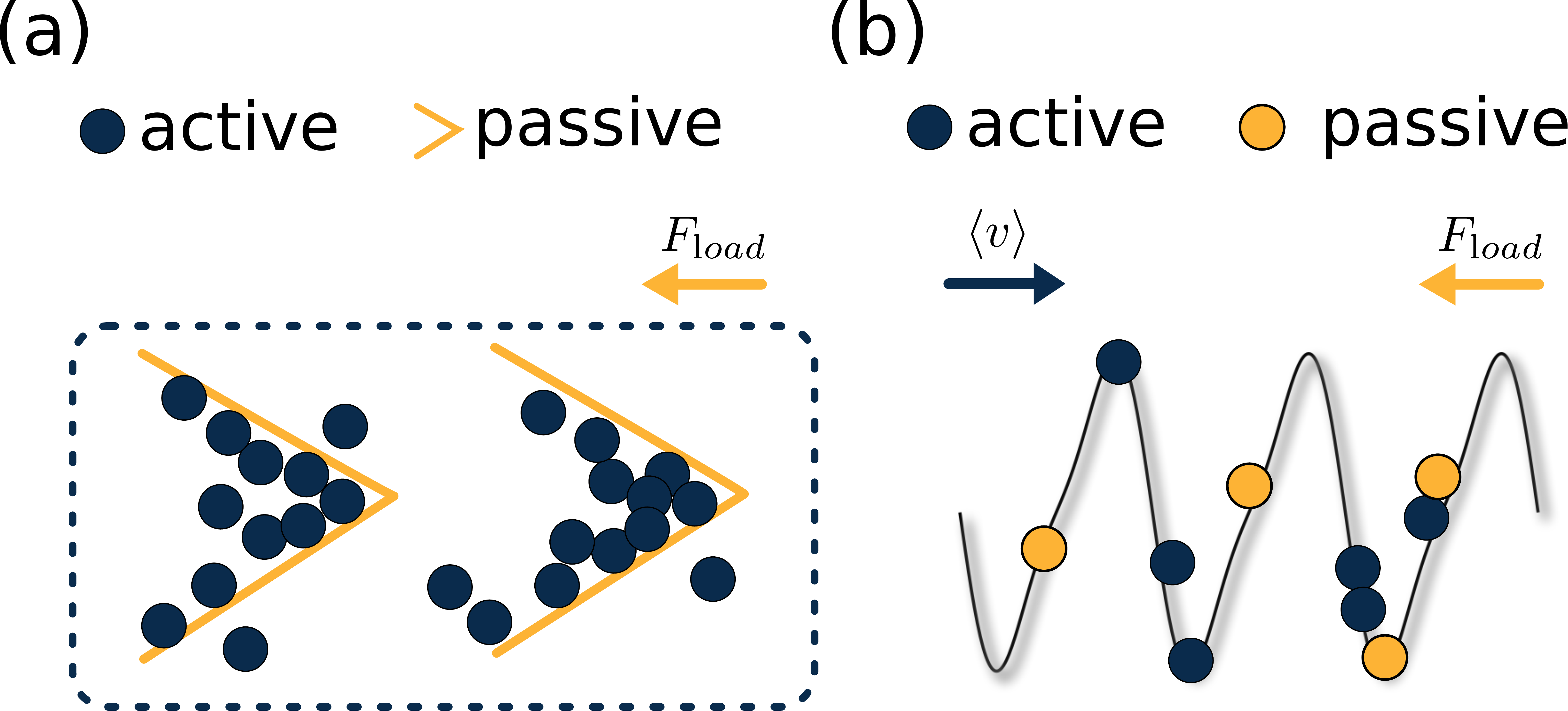}
\caption{ 
Different designs of active engines use different methods for breaking left-right symmetry. (a)~Schematic representation of a quasi-1D array where active particles push asymmetric ('v-shaped') passive obstacles against an external load, see for example \cite{pietzonka_autonomous_2019}. 
(b)~Schematic representation of a truly one-dimensional model where isotropic active and passive particles move subject to a sawtooth potential [Eq. \eqref{eq:sawtooth_pot}].  The active particles push the passive ones against an external load.}
\label{fig:scheme_ratchets}
\end{figure}

In this work, we consider a one-dimensional engine, built from an active/passive (AP) mixture.  Instead of pushing on localised obstacles, the particles feel forces from a ratchet potential, which acts throughout the system.  The interaction with this potential pushes the active particles to the right; the active particles push in turn on the passive particles, from which work can be extracted~\cite{fodor_active_2021}.  In contrast to engines based on asymmetric obstacles,  this design separates the spatial asymmetry (which is required for generating a current, and appears here via the ratchet potential) from the mechanism of work extraction (which happens through the isotropic passive particles).  Some aspects of these different design strategies are illustrated in Fig. \ref{fig:scheme_ratchets}.

In the following, we analyse a model similar to Fig. \ref{fig:scheme_ratchets}(b).  It consists of an AP mixture in one spatial dimension, where all particles are of equal size, interacting with an external sawtooth potential.  The particles interact by repulsive forces that are strong enough to prevent overtaking, so they move in single file.   In a previous work \cite{derivaux_rectification_2022-1}, we showed that the presence of active particles in this system generates a current of both active and passive ones.  To transform this system into an engine, we add a constant external force to all the passive particles, against which they do work.  We explore the principles that determine the efficiency of this simple design of bulk engine, with the aid of extensive numerical simulations.  We discuss how the insights from this minimal model can be applied in more general contexts.

The structure of the paper is as follows.  In Sec. \ref{sec:model}, we introduce {the equations of motion} for the AP mixture engine and define its efficiency. We will also describe the operation of the engine. In Sec. \ref{sec:opt_eff}, we study the engine's efficiency, including its dependence on model parameters. 
In particular, we show that the composition of the AP mixture strongly affects the efficiency, and we explain the mechanism for this.  
The single-file constraint means that the initial arrangement of active and passive particles is fixed by the initial condition and never changes: Sec. \ref{sec:sequences} discusses the effect of this initial arrangement on the results, and briefly addresses what happens when the constraint is relaxed.
Finally, conclusions and outlook are presented in Sec. \ref{sec:conclusions}.

\section{One-dimensional active engine}
\label{sec:model}

\subsection{System definition}
\label{subsec:model_def}

Following our previous work~\cite{derivaux_rectification_2022-1}, we consider a mixture of active and passive particles immersed in a 1D box of length $L$ with periodic boundary conditions. There are $N_a$ active run-and-tumble particles (RTPs) and $N_p$ passive particles.  Both active and passive particles feel external forces ($F_\mathrm{rat}$) which derive from a ratchet potential, as well as interaction forces ($F_{\rm int}$) that come from a Weeks-Chandler-Andersen (WCA) potential.  In addition, the passive particles feel a force $F_{\rm load}$ that represents an external loading, against which they do work.  This is illustrated schematically in Fig. \ref{fig:scheme_ratchets}(b).  The resulting model is very similar to that of~\cite{derivaux_rectification_2022-1}, the central difference being the loading force $F_{\rm load}$, which allows the system to do work on its environment. Imposing periodic boundary conditions is important to prevent particles building up on one side of the 1D channel due to rectification. Periodic boundary conditions represent an ideation of either a closed circular channel, ether a long microchannel constantly fed by large reservoirs of particles, conditions often met in real experimental devices \cite{locatelli_single-file_2016}.

Writing the position of the $i$th active particle as $x_{a,i}$ and that  of the $j$th passive particle as $x_{p,j}$, the equations of motion are 
\begin{align}
\dot{x}_{a,i} & = \frac{1}{\gamma} \left[ F_\mathrm{rat}({x}_{a,i}) + F_{\mathrm{int},i} \right] + v_0  \, \sigma_i (t) \label{eq:evo_active}  \\
\dot{x}_{p,j} & = \frac{1}{\gamma} \left[ F_\mathrm{rat}({x}_{p,j}) + F_{\mathrm{int},j} - F_\mathrm{load} \right] + \sqrt{2D_p} \, \xi_j (t) \;. \label{eq:evo_passive}
\end{align}
Here $\gamma^{-1}$ is the particle mobility (which is assumed to be equal for both species); $v_0$ is the self-propulsion speed of the RTPs, and $\sigma_i=\pm 1$ is their direction of propulsion, which flips randomly with rate $\alphaBare$.
Also $\xi_j$ is a unit Brownian noise acting on particle $j$, whose strength is set by the passive diffusion coefficient $D_p$.  Note that this noise is delta-correlated, 
$\langle \xi_i(t) \xi_j(t^{\prime}) \rangle = \delta_{ij} \delta(t-t^{\prime})$, while the direction of active self-propulsion is a telegraphic noise, correlated as $\langle \sigma_i(t) \sigma_j(t^{\prime}) \rangle = e^{-2\alphaBare \vert t-t^{\prime}\vert} \delta_{ij}$.  We assume that the white noises $\xi_j$ come from a thermal environment, so the corresponding bath temperature is $T=\gamma D_p/k_{\rm B}$.

The ratchet force is $F_\mathrm{rat}({x}) = -\nabla V(x)$ where $V$ is a sawtooth potential with wavevector $k$:
\begin{equation}
V(x) = - \frac{V_0}{2\pi} \left[ \text{sin}(2\pi kx) + \frac{1}{4}\text{sin}(4\pi kx)\right] \; ,
\label{eq:sawtooth_pot}
\end{equation}
as used in previous studies of current rectification~\cite{derivaux_rectification_2022-1,hanggi_artificial_2009}.  The force $F_{\rm int}$ derives from the interaction potential 
\begin{equation}
V_{\text{WCA}}\!\left( r \right) = 4\epsilon \left[ 
\left( \frac{\dBare}{r}\right)^{12} -  \left( \frac{\dBare}{r}\right)^6 + \frac14  \right] \theta\!\left(2^{1/6}\dBare - r \right)
\label{eq:WCA_potential}
\end{equation}
where $r$ is the distance between two particles, $\epsilon$ is the interaction strength, $d_0$ is the particle diameter, and $\theta$ is the Heaviside function.  All particles (both active and passive) interact via this same potential.%

We will use non-dimensional units to rescale and simplify the system parameters, as done in~\cite{derivaux_rectification_2022-1}.  Space units will be rescaled by the ratchet wavelength, $x^{\star} = k x$, and time units  by the time it takes a free active particle to translate one such wavelength $t^{\star} = k v_0 t$. Forces will be also expressed as multiple of the self-propulsion force of RTPs $\gamma_a v_0$. In these non-dimensional units, the equations of motion become
\begin{align}
\label{eq:EOM_a_nd} \frac{dx^{\star}_{a,i}}{dt^{\star}} & = F_\mathrm{rat}^{\star}(x^{\star}_{a,i}) +  F_{\mathrm{int},i}^{\star} + \sigma_i(t^\star)\\ 
\label{eq:EOM_p_nd} \frac{dx^{\star}_{p,j}}{dt^{\star}} & =  F_\mathrm{rat}^{\star}(x^{\star}_{p,j})  +  F_{\mathrm{int},j}^{\star} - F_\mathrm{load}^{\star}
 + \ell_{D} \, \xi_j^\star(t^\star)
\end{align}
in which
the non-dimensional forces are 
\begin{align}
F_\mathrm{rat}^{\star}(x^\star) & = f \left[ \text{cos}\left( 2 \pi x^{\star}\right) + \frac{1}{2} \text{cos}\left( 4 \pi x^{\star}\right) \right] \\
F_{\mathrm{int},i}^{\star} & = \epsilon_0 d \, \mathlarger{\sum}_{l (\neq i)} \left[ \left( \frac{d}{r^{\star}_{il}} \right)^{13} - \frac{1}{2}  \left( \frac{d}{r^{\star}_{il}} \right)^{7} \right] \theta\!\left(2^{1/6}d - \left\vert r^\star_{il}\right\vert \right) \\
F_\mathrm{load}^{\star} &= F_l \, ,
\end{align}
with rescaled parameters 
\begin{equation}
f = \frac{V_0 k}{\gamma v_0}\,, \qquad d = k\dBare\,, \qquad  \epsilon_0 = \frac{48 \epsilon }{\gamma v_0 k \left( \dBare\right)^{2} }\,, \qquad F_l= \frac{F_\mathrm{load}}{\gamma v_0} \,,
\label{eq:nondim_par1}
\end{equation}
and rescaled strengths for the random forces:
\begin{equation}
\alpha = \frac{\alphaBare}{k v_0}\,,  \qquad \ell_D = \sqrt{\frac{2D_pk}{v_0}} \;.
\end{equation}
Physically, $f$ is the strength of the ratchet force; $\epsilon_0$ is the strength of the interparticle repulsion; $F_l$ the magnitude of the external load; and $d$ is the ratio between the particle diameter and the wavelength of the ratchet potential.
Also $\alpha$ is the non-dimensional tumbling rate. 
The parameter $\ell_D$ is the ratio of a diffusive length to the ratchet wavelength, 
representing 
the root mean-squared displacement of a free passive particle, in the time for a free active particle to move one wavelength, without tumbling.
As well as the parameters appearing in the equations of motion, we also write 
\begin{equation}
N = N_a + N_p, \qquad \phi_a = \frac{N_a}{N}, \qquad c = \frac{N}{k L} \,,
\end{equation} 
which are the total number of particles, the fraction of active particles, and the total (dimensionless) concentration, respectively.

Together, these 8 parameters $\left(f,d,\epsilon_0,F_l,\alpha, \ell_D, \phi_a, c \right)$ fully describe the AP mixture and the forces present in the system, while $N$ sets the system size. 
In the rest of the paper, we only consider the non-dimensionalised {equations of motion} and we drop the star notation used in Eqs. \eqref{eq:EOM_a_nd} and \eqref{eq:EOM_p_nd}.

\subsection{Steady state averages and the role of initial conditions}

We analyse the steady state of this non-equilibrium system.  An important quantity is the average velocity of the active particles, which can be estimated as 
\begin{equation}
\langle v_{a} \rangle \approx 1/(N_a\tau) \sum_{i=1}^{N_a} \int_{t_{\rm eq}}^{t_{\rm eq}+\tau} \dot{x}_{a,i}(t') dt' \,,
\label{equ:va-ave}
\end{equation}
 where the times $\tau$ and $t_{\rm eq}$ are respectively chosen large enough that effects of initial transients can be neglected, and statistical uncertainty is not too large to affect the results.
The WCA interaction potential in this model means that particles can never pass each other.  The resulting single-file motion means that both passive and active particles must have the same average velocity, $\langle v_{p} \rangle = \langle v_{a} \rangle$: this is the \emph{ratchet current}, which we denote simply by $\langle v\rangle$.

However, the single-file constraint has an additional effect, which is that the ordering of the active and passive particles is preserved for all time. As a result, steady state averages (including $\langle v \rangle$~\cite{derivaux_rectification_2022-1}) may depend on the initial placement of the particles.  This dependence is discussed in Sec.~\ref{sec:sequences}.  Unless explicitly stated, all steady state averages in this work are computed by averaging over long time periods and over several (randomly chosen) initial particle sequences.  In particular, the steady state average velocity $\langle v\rangle$ is obtained by a three-level average: first as a time average as in \eqref{equ:va-ave}, then over all particles (of both species), and finally over several independent simulations with different initial particle sequences. 
In the following, angle brackets always represent such three-level averages.

\begin{figure}
\includegraphics[width = 1.00 \textwidth]{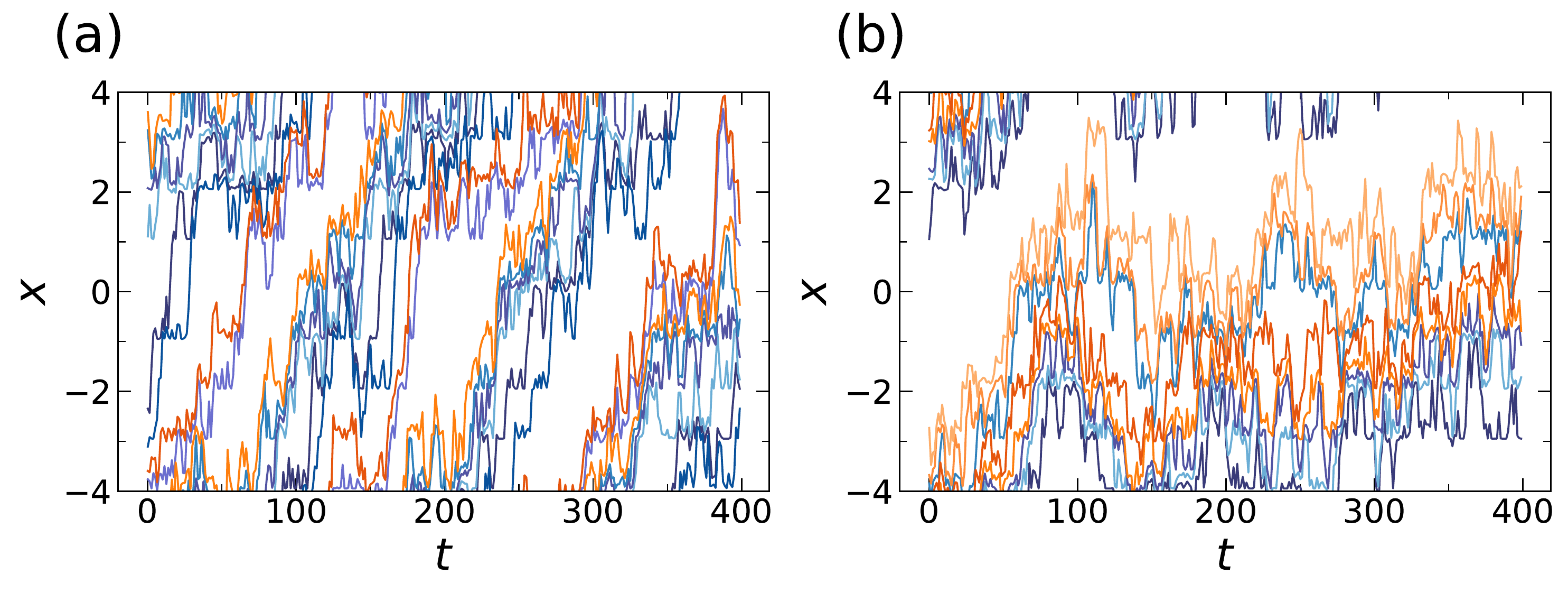}
\caption{Typical trajectories of the AP mixture working as an engine for (a) a mixture rich in active particles ($\phi_a = 0.75$) and (b) a half-half mixture  ($\phi_a = 0.5$). Trajectories of RTPs are in shades of blue and those of Brownian particles in shades of orange. Amplitude of the external load is $F_l =0.1$ in both cases. \\
Parameters: $N = 8$, $F_l = 0.1$, $f =0.75$, $\alpha = 1$, $\ell_D=0.5$, $\epsilon_{0} = 0.25$, $d = 0.2$, $c=1$, $\tau =400$, $t_{\text{eq}} = 200$, $dt = 2.5 \times 10^{-6}$. }
\label{fig:tseries_comp}
\end{figure}

\subsection{Operation of the system as an active engine}
\label{subsec:operation}

For $F_l>0$, the system defined here behaves as an active engine.  To see this, consider first the system without any load ($F_l=0$).  The active particles are constantly doing work, in order to self-propel themselves. Together with the ratchet potential and the particle interactions, this work generates a non-zero average steady-state velocity $\langle v\rangle$.  For the parameters considered here, $\langle v\rangle>0$.  In this case, the system operates effectively as a rectification device, but all the work done by the active particles is dissipated as heat.  

When a load is added ($F_l>0$), the passive particles do work against this force at rate
\begin{equation}
P =  N_p  F_l \langle v \rangle  \, .
\label{eq:power_def}
\end{equation}  
That is, the rectification effect can be used to convert the undirected work of the individual particles into a coherent effort that acts against an applied load. Note that $P$ is extensive in the number of passive particles, since they all feel the same loading force. Sample trajectories for this case are shown in Fig.~\ref{fig:tseries_comp}.

To analyse the operation of such an engine, we compare the rate of extracted work (output power, $P$) with the rate at which the underlying active particles do work. The rate of active work for a single RTP is $\sigma_i(t) \dot{x}_{a,i}(t)$, so the average rate of work for all such particles in the steady state is
\begin{equation}
\dot{W}_a  =   \sum_{i=1}^{N_a} \langle \sigma_i \dot{x}_{a,i}  \rangle \,.
\label{equ:Wa_defn}
\end{equation} 
An efficiency of this engine is then defined as the ratio of the output power $P$ to the rate of active work for the RTPs:
\begin{equation}
\mathcal{E} =  \frac{P}{\dot{W}_a} \,.
\label{eq:eff_definition}
\end{equation}
We emphasize again that $P$ is obtained by summing over all passive particles, and $\dot{W}_a$ is summed over all active particles. This point will be important below, when analysing how the efficiency varies as one changes the composition of the mixture.  

It is also important that the efficiency $\mathcal{E}$ is a ratio of two work rates, in contrast to the usual situation for heat engines, where efficiency is the ratio of the work output to the heat input.  Depending on the physical mechanism by which the active particles propel themselves, one might also consider the ratio between $\dot{W}_a$ and the rate with which the particles consume their individual fuel sources.  This ratio feeds into the practical efficiency of active engines~\cite{pietzonka_autonomous_2019,parmeggiani_energy_1999}, but we do not consider it here. (The fuel consumption clearly involves more microscopic information than we used to define the RTP dynamics of our model.) 

\begin{figure}
\includegraphics[width = 1.00 \textwidth]{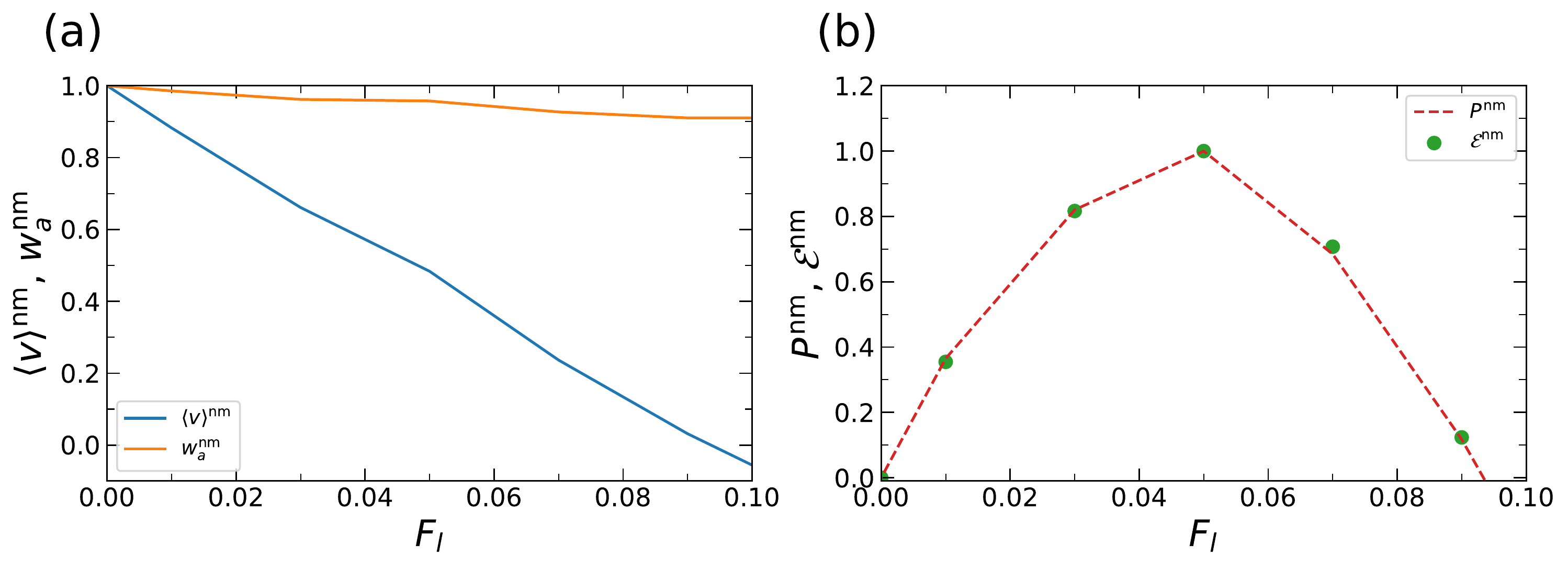}
\caption{(a)~Dependence of ratchet current and active work on external passive load $F_l$.  The ordinates are normalised by their respective values at $F_l = 0$: normalised current $\langle v\rangle^{\mathrm{nm}} = \langle v\rangle / \langle v\rangle_0$ and normalised active work $w_a^{\mathrm{nm}}=   w_a / \langle w_a \rangle_0$.  (b) Dependence of power output $P$ and efficiency $\mathcal{E}$ on $F_l$. Power and efficiency are normalised according to their respective maximal value when $F_l$ is varied:  $P^{\mathrm{nm}} = P/\, \text{max}_{F_l} P$ and $\mathcal{E}^{\mathrm{nm}} = \mathcal{E}/ (\text{max}_{F_l} \mathcal{E})$. \\
Parameters are those of Eq.\eqref{equ:baseline}. }
\label{fig:panel_currentactiwork_powereff_Fext_64part_phia05}
\end{figure}

To illustrate these ideas, we present simulation results in Fig. \ref{fig:panel_currentactiwork_powereff_Fext_64part_phia05}.  For a representative set of parameters with $N_a=N_p$, we increase the load $F_l$, plotting the mean velocity of the system, as well as the mean active work per (active) particle
\begin{equation}
w_a = \frac{\dot{W}_a}{N_a} \, ,
\label{equ:wa_defn}
\end{equation}
which is an intensive quantity.

To see clearly the dependence of different quantities on the load $F_l$, we write $\langle v \rangle_0$ and $\langle w_a\rangle_0$ for the average velocity and active work in the absence of any load.  
Since the applied load acts against the ratchet current, one expects it to reduce the ratchet velocity $\langle v \rangle$. Fig.~\ref{fig:panel_currentactiwork_powereff_Fext_64part_phia05}(a) confirms this expectation, showing that the reduction is well-described by a linear relationship: 
\begin{equation}
\langle v \rangle \approx \langle v \rangle_0 - \chi F_l
\label{equ:v-chi}
\end{equation} 
 where $\chi$ is a response coefficient.  Hence by \eqref{eq:power_def}, 
$P \approx F_l [  \langle v \rangle_0 - \chi F_l ]$
has a quadratic dependence on $F_l$, as confirmed in Fig.~\ref{fig:panel_currentactiwork_powereff_Fext_64part_phia05}(b).  

Fig.~\ref{fig:panel_currentactiwork_powereff_Fext_64part_phia05}(a) shows that the dependence of $w_a$ on $F_l$ is much weaker than that of $\langle v \rangle$, similar to other active engines~\cite{fodor_active_2021,pietzonka_autonomous_2019}.  
Fig.~\ref{fig:panel_currentactiwork_powereff_Fext_64part_phia05}(b) shows the efficiency and power, again normalised by their values at $F_l=0$.  {Since $\dot{W_a}$ can be approximated as independent of $F_l$ in \eqref{eq:eff_definition}},  one sees that the efficiency $\mathcal{E}$ follows the same trend as the power $P$, with their maxima appearing at the same load $F_l$.  In this case there is no trade-off between power and efficiency, in contrast to classical (cyclic) thermal heat engines.  (In those systems, the denominator that appears in the efficiency usually increases significantly as one increases the cycle frequency, in which case maximal efficiency occurs at minimal power output.)

\section{Efficiency optimisation}
\label{sec:opt_eff}

Having illustrated the operation of the engine, we now give a more detailed analysis of its dependence on the various system parameters.  We aim to identify parameters leading to high efficiency $\mathcal{E}$, and  characterise the associated physical behaviour.

As baseline parameters, we take 
\begin{equation}
\begin{split}
(f,\alpha,\ell_D,d,c,\epsilon_0)&=(0.75,1,0.5,0.2,1,0.25) \\
(\tau, t_{\rm eq}, dt) &= (800, 200, 2.5 \times 10^{-6})
 \, ,
 \end{split}
\label{equ:baseline}
\end{equation}
 as in Fig.~\ref{fig:panel_currentactiwork_powereff_Fext_64part_phia05}.   These values are representative of the regime where a significant current is generated in a system without load, and they also lead to an effective engine.  We fix $f$ and $\epsilon_0$ throughout and we consider effects of varying all other parameters. (In justification of this choice, note that the behaviour depends only weakly on $\epsilon_0$.  Dependence of the ratchet current on $f$ was discussed extensively in~\cite{derivaux_rectification_2022-1}, we expect the results shown here to be qualitatively similar whenever $f$ is in the range of effective ratchet operation.)

\subsection{The role of composition}
\label{sec:composition}

\begin{figure}
\includegraphics[width = 0.95 \textwidth]{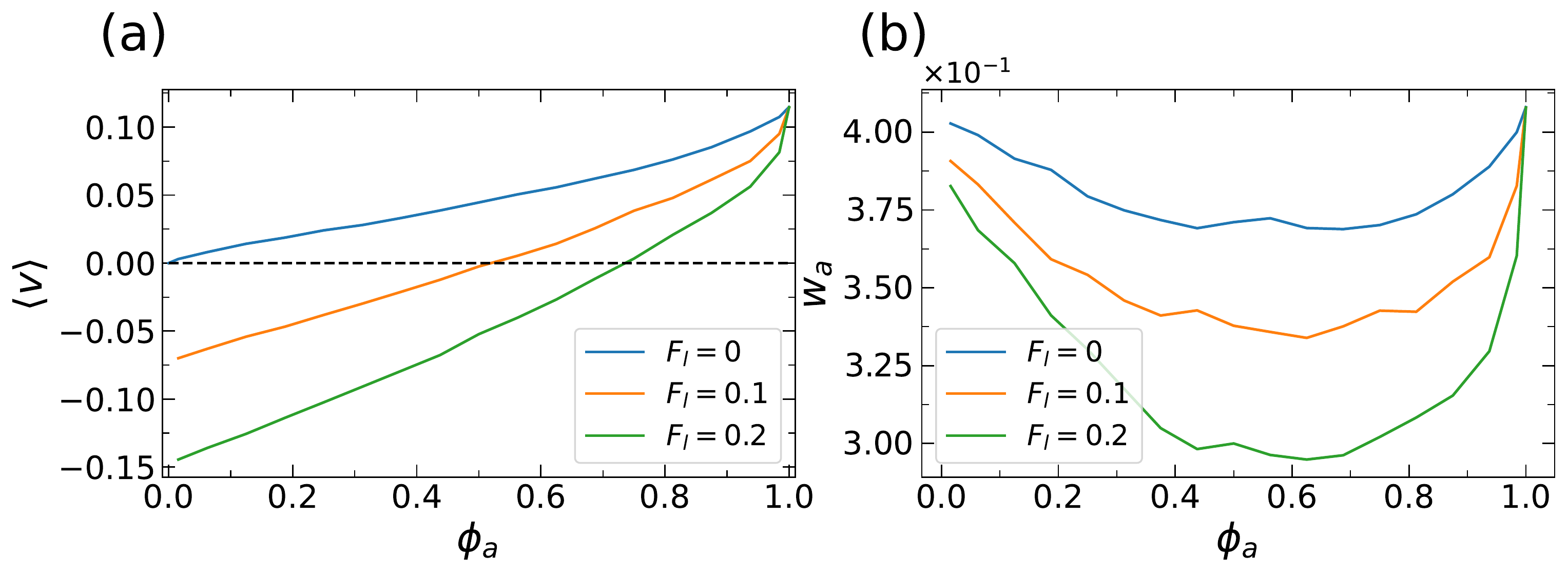}
\caption{Dependence on composition of (a) current per particle and (b) active work per active particle, for various values of $F_l$.
Parameters: baseline Eq.\eqref{equ:baseline} and $N = 64$. }
\label{fig:current_activework_scaling_intensive_64part}
\end{figure}

The composition of the AP mixture strongly affects  the behaviour of the engine, and its efficiency.  Fig. \ref{fig:current_activework_scaling_intensive_64part} shows the current per particle $\langle v\rangle$ and the active work, as the fraction of active particles $\phi_a$ is increased.  The ratchet velocity generally increases with $\phi_a$: passive particles act as obstacles which hamper the current, so reducing the passive fraction by raising $\phi_a$ always increases $\langle v \rangle$.

In the absence of any load, Fig. \ref{fig:current_activework_scaling_intensive_64part}(a) shows that the velocity is approximately linear in $\phi_a$ over a wide range of compositions, with some deviations near $\phi_a=1$.  However, Fig. \ref{fig:current_activework_scaling_intensive_64part}(b) shows that the (intensive) active work per particle $w_a$ is non-monotonic in $\phi_a$ with a minimum at $\phi_a\approx 0.5$.  Using  \eqref{equ:Wa_defn} with (\ref{eq:EOM_a_nd},\ref{equ:wa_defn}), one sees that
$
w_a =  v_0^2 - \langle  \sigma_i [ F_\mathrm{rat}(x_i) +  F_{\mathrm{int},i}] \rangle
$.
The interaction term $\sigma_i F_{\mathrm{int},i}$ leads to a significant reduction of $w_a$ when particles self-propulsion is blocked by other particles.   Active particles alone tend to have a positive velocity (due to rectification) while isolated passive ones diffuse symmetrically.  Hence, the passive particles will often block the active ones, leading to positive values of $\langle \sigma_i F_{\mathrm{int},i}\rangle$ and suppressing $w_a$.  Since this effect relies on active-passive collisions, it is expected to be strongest for $\phi_a\approx 0.5$, as observed.

\begin{figure}
\includegraphics[width = 1.0 \textwidth]{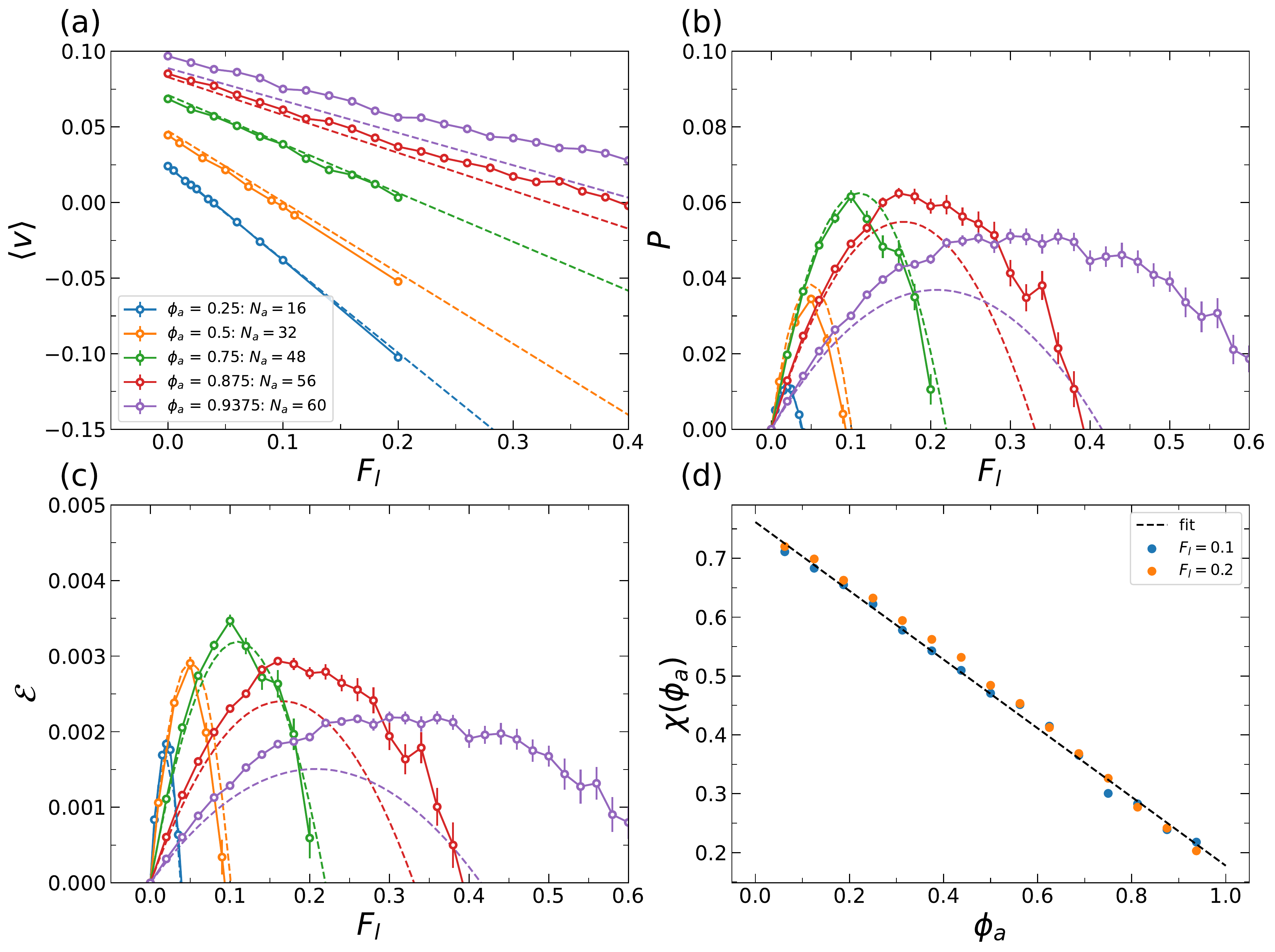}
\caption{Dependence on $(\phi_a, F_l)$ of different system quantities: (a) the ratchet current $\langle v \rangle$, (b) the power output $P$, (c) the efficiency $\mathcal{E}$ and (d) the response coefficient $\chi$. Circles represent results from numerical simulations, with error bars of $\pm 1 \text{SEM}$. Dotted lines are fits to \eqref{eq:v-model} using linear regression. Parameters: same as Fig. \ref{fig:current_activework_scaling_intensive_64part}. }
\label{fig:theo_fit_4panels_EB}
\end{figure}

To capture the effects of composition, we computed the average velocity as well as the power and efficiency of the engine as a function of load $F_l$, for various $\phi_a$.  Results are shown in Fig. \ref{fig:theo_fit_4panels_EB}, for a system of 64 particles.  When comparing these results with Fig.~\ref{fig:current_activework_scaling_intensive_64part}, note that the efficiency $\mathcal{E}$ in \eqref{eq:eff_definition} can be expressed as
\begin{equation}
\mathcal{E} = \frac{F_l \langle v \rangle}{ w_a } \times \frac{1- \phi_a}{\phi_a} \, .
\label{equ:e-phi}
\end{equation}
The second factor in this expression reflects the fact that the power $P$ is extracted from passive particles while the active work comes from the active ones.   For high efficiency, $\phi_a$ should not be too large, otherwise there are too few passive particles from which work can be extracted; neither should $\phi_a$ be too small, since that leads to $\langle v \rangle<0$ [recall Fig.~\ref{fig:current_activework_scaling_intensive_64part}(a)], where no power can be extracted.

To make further progress, note that \eqref{equ:e-phi} depends on $\phi_a$ explicitly, and also implicitly via $\langle v \rangle$ and $w_a$.  
To unpack the behaviour of $\langle v\rangle$, we show in Fig. \ref{fig:theo_fit_4panels_EB}(a) its dependence on both $F_l$ and $\phi_a$.   The results are consistent with \eqref{equ:v-chi}, with a $\phi_a$-dependent response coefficient $\chi$, shown in Fig.~\ref{fig:theo_fit_4panels_EB}(d).  On increasing $\phi_a$, the response $\chi$ to the load $F_l$ is reduced; the effect can be modelled as
$\chi(\phi_a) = p + q \, (1-\phi_a)$ where $(p,q)$ are fitting parameters.  (Recall that all passive particles feel the same load force $F_l$, so reducing the number of passive particles reduces the \emph{total} load on the system, and it is natural that $\chi$ should decrease.)

Combining the approximate linear scaling of $\langle v \rangle_0$ with $\phi_a$ [Fig.~\ref{fig:current_activework_scaling_intensive_64part}(a)] and the fit for $\chi$,
we obtain the following approximate formula for the ratchet current:
\begin{equation}
\langle v \rangle \approx r \phi_a - [ p + q (1-\phi_a) ] F_l  \;,
\label{eq:v-model}
\end{equation}
which now depends on three fitting parameters $r$, $p$ and $q$.  Those three parameters were estimated through a linear regression the data in Fig. \ref{fig:theo_fit_4panels_EB}(a).  We estimate
$\left( r, p, q \right) = \left( 0.095, 0.17, 0.58 \right)$.

In comparison to $\langle v\rangle$, the active work $w_a$ depends relatively weakly on $\phi_a$ [Fig.~\ref{fig:current_activework_scaling_intensive_64part}(b)].  As a first approximation, we neglect this dependence, and use \eqref{eq:v-model} to predict the $\phi_a$-dependent power and efficiency via \eqref{eq:power_def} and \eqref{equ:e-phi}.  These predictions are shown in Fig. \ref{fig:theo_fit_4panels_EB}(b,c): there is reasonable agreement with the data, especially given the simple linear dependences that appear in \eqref{eq:v-model}.  In  Appendix \ref{app:activework} we analyse a more refined theory where the $\phi_a$-dependence of $w_a$ is taken into account; the results are broadly similar.

Fig. \ref{fig:theo_fit_4panels_EB} does show significant differences between the approximate predictions of (\ref{eq:v-model}) and the data when $\phi_a$ is large.  In this case, (\ref{eq:v-model})  underestimates the efficiency.  (We found that this fit is not improved by taking into account the $\phi_a$-dependence of $w_a$.)  We also recall the deviations from linearity at large $\phi_a$ in Fig.~\ref{fig:current_activework_scaling_intensive_64part}(a), and we note that the data for $\chi(\phi_a)$ extrapolate to a finite value ($p$) as $\phi_a\to1$, but for $\phi_a=1$ there are no passive particles and the load $F_l$ has no effect, hence one should have $\chi=0$ there.  These observations point to interesting behaviour for large $\phi_a$, which we discuss further below.

\subsection{Optimising the efficiency through composition change}

This Section discusses how the performance of the engine can be optimised by adjusting $\phi_a$, based on the results obtained so far.
For a fair comparison, we show data after optimising the load force $F_l$ (keeping all other variables fixed).  The physical interpretation of this optimisation is simple because substituting \eqref{equ:v-chi} into \eqref{equ:e-phi}, we see that $\mathcal{E}$ has a quadratic dependence on $F_l$ and hence
\begin{equation}
\mathcal{E}^* \equiv \max_{F_l} \mathcal{E} \approx \frac{\langle v \rangle_0^2(1- \phi_a)}{ 4\chi w_a \phi_a}
\label{equ:emax_approx}
\end{equation}
with the maximum occurring at $F_l^*\approx \langle v \rangle_0/(2\chi)$.  That is, if the ratchet current depends weakly on $F_l$ then $\chi$ is small, efficient performance occurs for large loads, and this also tends to result in high efficiency.

\begin{figure}
\includegraphics[width = 0.95 \textwidth]{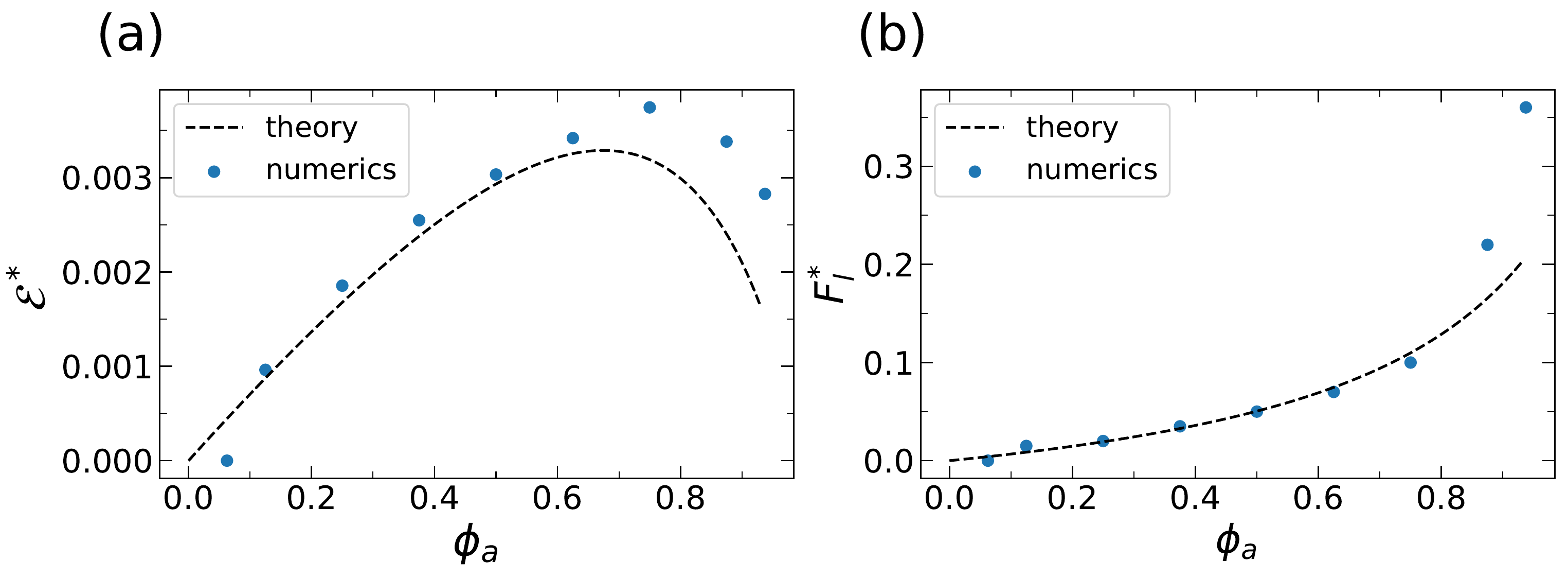}
\caption{Plots of (a) the maximum efficiency of the engine and (b) of the external force $F_l$ at maximum efficiency when active particles fraction is raised. Dashed lines represent the theoretical predictions coming from Eq. \eqref{equ:emax_approx_scalings}. Parameters are the same as Fig.~\ref{fig:current_activework_scaling_intensive_64part}. }
\label{fig:maxe_theovsnum_panel}
\end{figure}

Inserting the fitted $\phi_a$-dependence of $\langle v\rangle_0$ and $\chi$ from above, this becomes
\begin{equation}
\mathcal{E}^* \approx \frac{r^2\, \phi_a \, (1- \phi_a)}{ 4w_a \left[p + q(1-\phi_a)\right] }\,.
\label{equ:emax_approx_scalings}
\end{equation}
We see that dependence of efficiency on $\phi_a$ has two regimes: for small $\phi_a$ it increases linearly from zero as active particles are introduced and start to do work; for large $\phi_a$ (close to 1) it decreases again because there are few passive particles from which work can be extracted.  
This non-monotonic evolution is consistent with the numerical simulations shown in Fig. \ref{fig:maxe_theovsnum_panel}(a), where the dashed line represents the theoretical prediction of Eq.~\eqref{equ:emax_approx_scalings}. They capture the maximal efficiency obtained for a broad range of $\phi_a$ up to $\phi_a = 0.75$.  For larger $\phi_a$, the qualitative behaviour is captured by the theory, but there are significant deviations: Eq.~\ref{eq:v-model} underestimates the efficiency.
Fig.~\ref{fig:maxe_theovsnum_panel}(b) shows that for large $\phi_a$, the efficient operation of the engine relies on increasingly large load forces $F_l$.

Note that for $\phi_a=1$ there are no passive particles on which the load is acting, and the velocity $\langle v\rangle$ is independent of $F_l$.  From \eqref{eq:v-model}, this means that $\chi=0$ for $\phi_a=1$.  However, extrapolating the data of Fig.~\ref{fig:theo_fit_4panels_EB}(d) towards $\phi_a=1$, the behaviour is not consistent with this limit: it seems that $\chi$ tends to $p\approx 0.17$.  Indeed, the limit $\phi_a\to1$ must be somewhat singular because the behaviour with no passive particles at all is qualitatively different to the behaviour with a single passive particle on which a load is acting.  Still, it is interesting to note that if $p=0$ then \eqref{equ:emax_approx_scalings} would predict maximal efficiency as $\phi_a\to1$ (corresponding to a single passive particle in a system with many active ones).  Otherwise the optimal efficiency occurs for $\phi_a<1$: the smaller is $p$, the larger is the optimal $\phi_a$.  In practice $p>0$ and the optimal efficiency is at $\phi_a<1$: efficient operation involves multiple passive particles against which the active particles do work.

\begin{figure}
\includegraphics[width = 0.66 \textwidth]{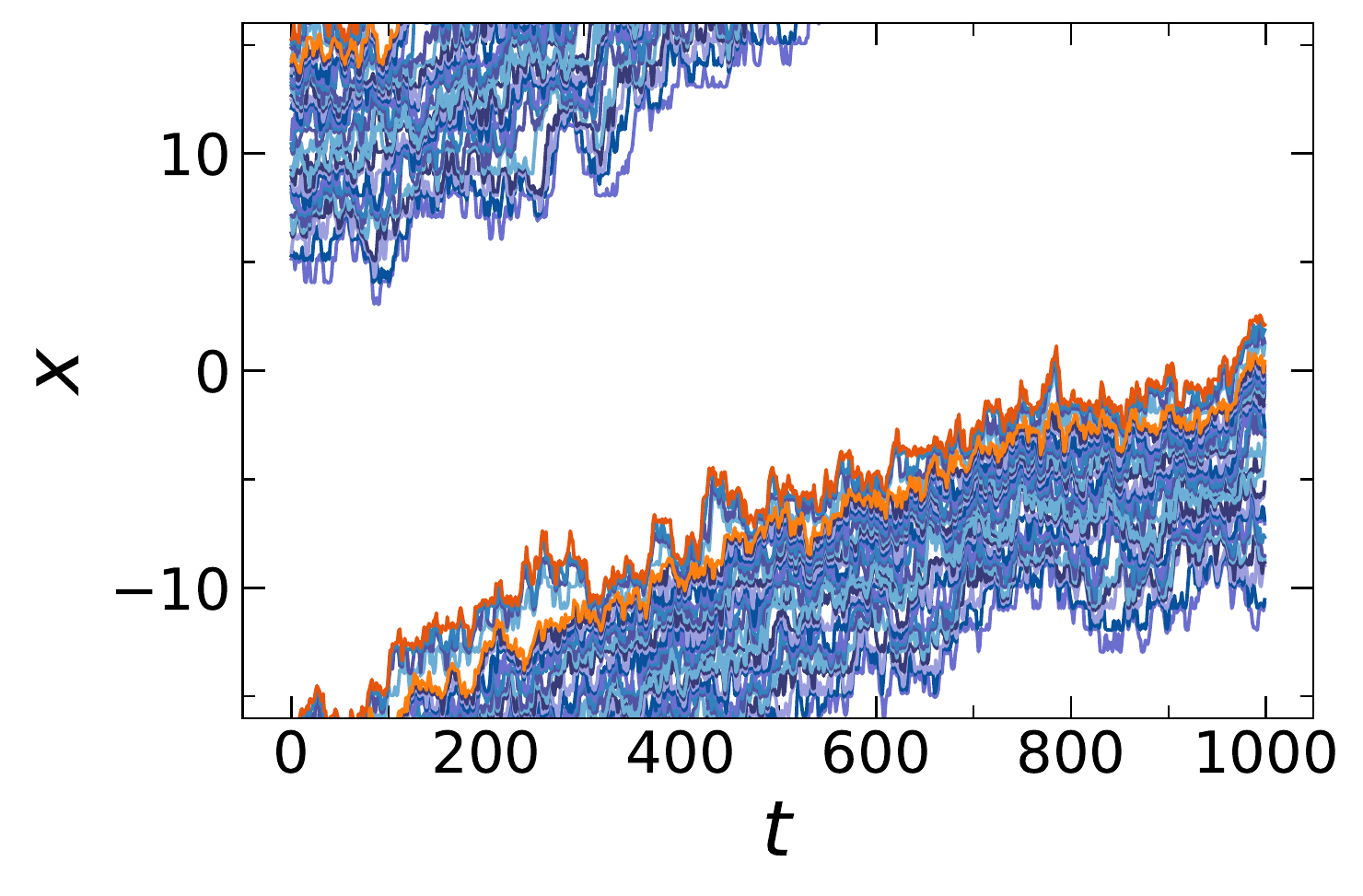}
\caption{Illustration of a representative trajectory in which large teams of active particles are pushing a few passive particles with a particular high load (here $F_l = 0.5$). Shades of blue denote active particles trajectories and shades of orange passive ones. \\
Parameters: baseline \eqref{equ:baseline} and $N = 32$, $\phi_a = 0.9375$, $F_l = 0.5$, $\tau =1000$, $t_{\text{eq}} = 1000$.}
\label{fig:tseries_32part_collective}
\end{figure}

An interesting feature of Fig. \ref{fig:maxe_theovsnum_panel}(a) is that the optimal efficiency significantly exceeds the prediction of the approximate theory based on \eqref{equ:emax_approx_scalings}, suggesting that deviations from this simple picture at large $\phi_a$ are contributing positively to the operation of the engine.  To illustrate the physical origin of this effect, Fig.~\ref{fig:tseries_32part_collective} shows a trajectory of the system at large $\phi_a= 1/16$.  The physical picture is that a passive particle with a large load $F_l$ is being pushed along by a large ``team'' of active ones.  

This team is a result of self-organisation in this non-equilibrium system.  It is not surprising that such non-equilibrium steady states -- which also correspond to large loads $F_l$ -- cannot be described by the simple linear theory.  As a simple model for this regime, one might consider a system with just one passive particle, which acts as a moveable wall against which all particles are pushing.  This would lead to connections with theories of active pressure \cite{takatori_swim_2014,takatori_towards_2015,solon_pressure_2015,solon_pressure_2015-1,winkler_virial_2015,nikola_active_2016,speck_ideal_2016}. However, we focus here on systems with finite densities of passive particles, since the optimal efficiency lies in that regime.  Since the optimal engine performance occurs when $\phi_a$ is significantly bigger than $0.5$, such states do involve teams of active particles pushing on a smaller number of passive ones, although the situation is less extreme than that of Fig.~\ref{fig:tseries_32part_collective}.

\subsection{Effect of geometric parameters}

\begin{figure}
\includegraphics[width = \textwidth]{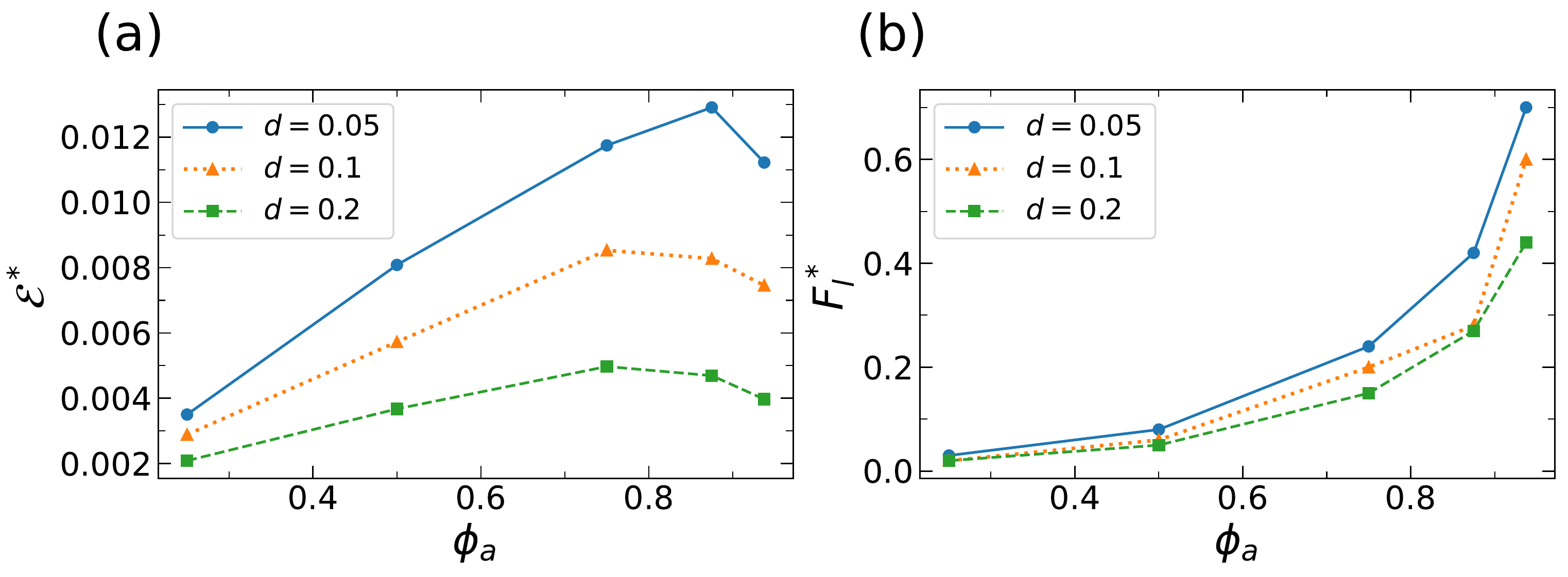}
\caption{Effect of varying $d$.  (a)~For any composition $\phi_a$, maximal efficiency is favoured by reducing the particle size, relative to the the wavelength of the potential. (b) Reducing the particle size also increases the optimal load $F_l$.\\
Parameters: baseline Eq.~\eqref{equ:baseline} with variation of $d$ as shown, also $N = 16$. }
\label{fig:size_panel}
\end{figure}

As well as the composition, the performance of the engine also depends on 
the total particle concentration $c$, and 
the ratio between the particle size and the wavelength of the sawtooth potential [defined as $d$ in \eqref{eq:nondim_par1}]. 
To analyse this,
we vary the model parameters $(\phi_a,d,c)$: for each set of parameters we optimise the load $F_l$ to find the maximum efficiency.
Fig.~\ref{fig:size_panel} shows the resulting efficiencies as $d$ is varied, together with the load forces that achieve them.  
The main observation is that smaller particles lead to higher efficiency, which is associated with (slightly) higher loads.
It was already observed in \cite{derivaux_rectification_2022-1} that smaller particles lead to higher ratchet currents, so this is not surprising.  This may be partly attributable to smaller active particles having fewer collisions, which tend to reduce the ratchet efficiency.  

In the situation where many active particles push together as a team against a single passive obstacle, an additional consideration is the number of particles that can fit in a single minimum of the potential: the collective pushing effect should be more efficient when the active particles are all in the same potential well.  Overall, it seems that reduction in $d$ is a potentially useful mechanism for increasing efficiency, especially because this parameter should be relatively easy to measure and control in practical settings, {\em e.g.}, by tuning the ratchet potential wavelength.

\begin{figure}
\includegraphics[width = \textwidth]{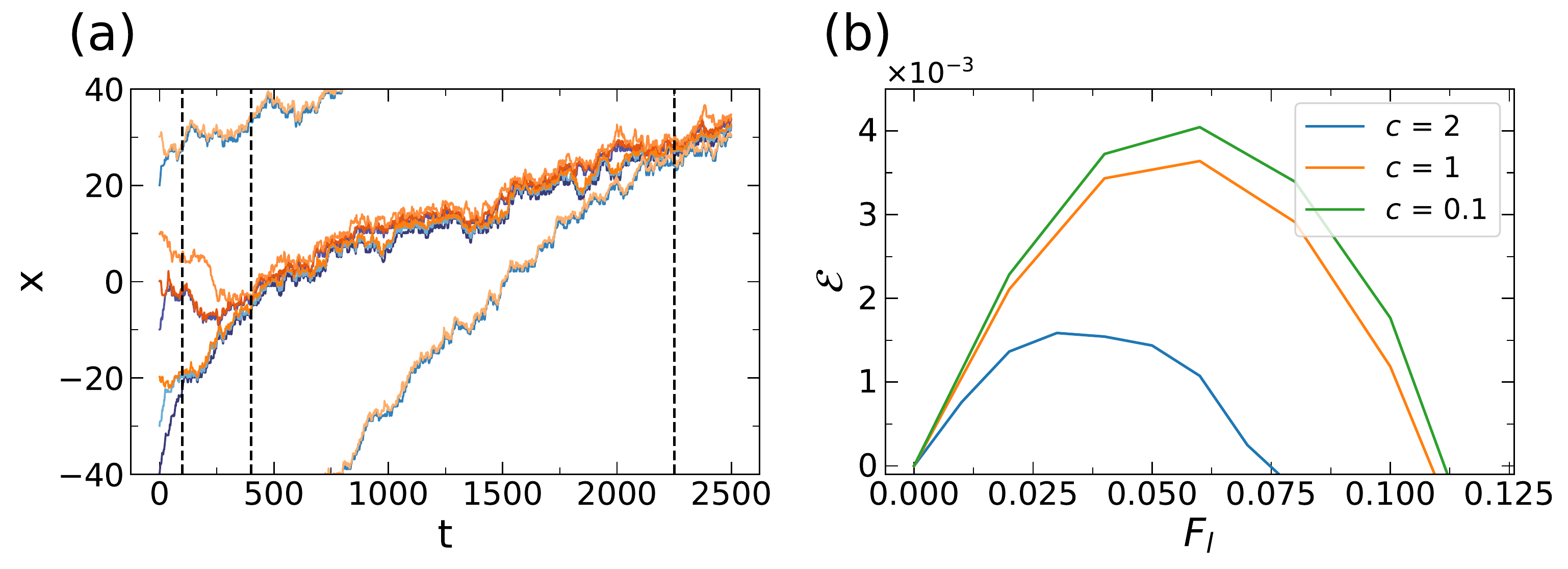}
\caption{(a) Snapshot of particle trajectories in a dilute system, $c=0.1$. Shades of blue represent RTPs and shades of orange Brownian particles. (b) Effect of dilution on plots of efficiency against external drive. \\
Parameters: (a) $N = 8$, $F_l = 0.1$, $\tau = 2500$, $t_{\text{eq}} = 0$, (b) $N = 16$, $\tau = 800$, $t_{\text{eq}} = 200$ for $c = 1,2$ and $\tau = 2000$, $t_{\text{eq}} = 1000$ for $c = 0.1$. Common parameters to both figures: baseline Eq.~\eqref{equ:baseline}. }
\label{fig:panel_tseries_conc}
\end{figure}

As an alternative method for avoiding particle collisions, we performed simulations with reduced concentration $c$.  For the representative case of $\phi_a=0.5$, Fig. \ref{fig:panel_tseries_conc}(b) shows that dilution tends to increase the efficiency, but the effect saturates quickly for $c\lesssim 1$ (corresponding to fewer than one particle per wavelength of the potential).
The mechanism for this saturation is illustrated by the representative dynamical trajectory in Fig.~\ref{fig:panel_tseries_conc}(a), in which blue lines are trajectories of active particles, while evolution of passive particles forms the orange lines. 
We see small teams of active particles that push passive ones; as time progresses then the resulting clusters of particles tend to merge, with several passive particles being pushed by a larger number of active ones.
Dashed bars show the times at which clusters are formed. Around $t=100$, the active particles self-organise into teams, and the resulting clusters of particles start  merging around $t=400$. All these clusters eventually merge into a single large cluster around $t=2200$. Note that the average speed always decreases after each clustering event. In the long time limit, all the active and passive particles will presumably form a large localised cluster.  At this point the overall concentration is irrelevant because all the particles are localised in a single part of the system, hence the efficiency becomes independent of $c$.

\subsection{Effect of particle dynamics on efficiency}

We now consider effects of the RTP tumble rate $\alpha$ and the passive particle diffusivity (parameterised by $\ell_D$).  In the absence of any load, these parameters have significant effects on the ratchet velocity.  For example, it was shown in~\cite{derivaux_rectification_2022-1} that rectification is effective when $\alpha$ and $\ell_D$ are both small (assuming that other parameters are chosen appropriately).  This corresponds to a large persistence length for the active particles, aiding rectification, and weak thermal noise for the passive ones (otherwise these noises can disrupt the rectification effect).

For the active engine, we note that efficiency depends strongly on the load force $F_l$, as well as on the composition. {While the previous Section was focused on the exploration of the geometric parameter space $(F_l,\phi_a,d,c)$, we now perform simulations that explore the four-dimensional parameter space spanned by $(F_l,\phi_a,\alpha,\ell_D)$, corresponding to the different noise types of particles dynamics.}  To summarise the resulting data, we again extract the maximal efficiency for each combination of $\left( \alpha, \ell_D, \phi_a \right)$, by first optimising over the load force $F_l$.

\begin{figure}
\includegraphics[width = \textwidth]{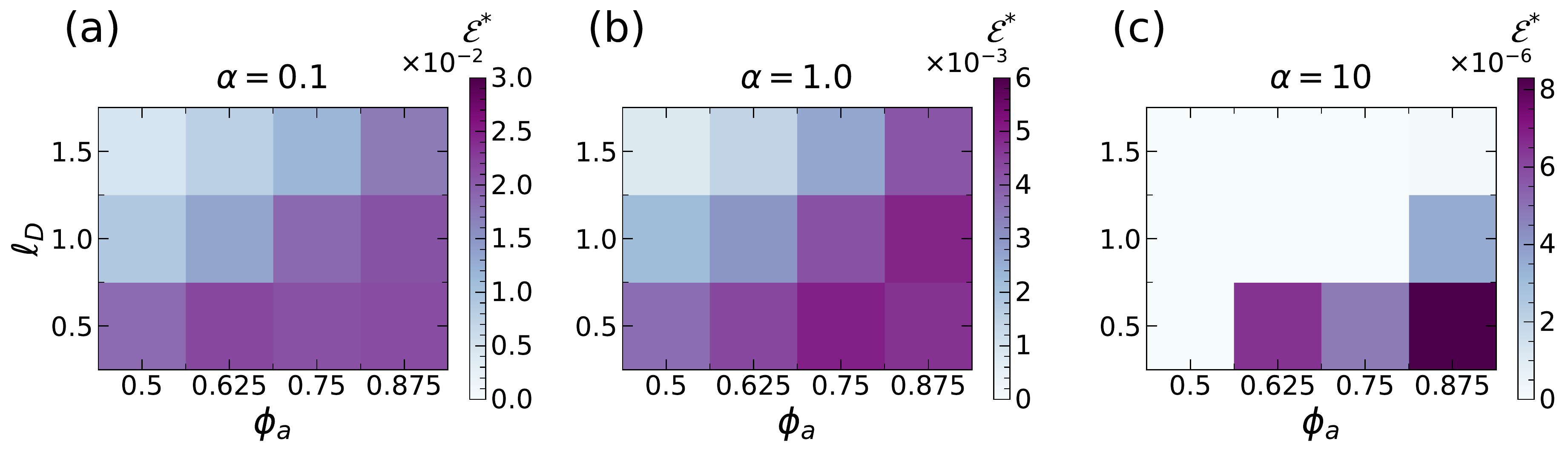}
\caption{Heat maps representing the maximal efficiency for each set of $\left( \alpha, \ell_D, \phi_a \right)$. Notice that the color map range is different from each different value of $\alpha$.\\
Parameters: baseline \eqref{equ:baseline} with variation of $\alpha$ and $\ell_D$, $N = 16$. }
\label{fig:panel_alpha_elld}
\end{figure}

Results are plotted in Fig.~\ref{fig:panel_alpha_elld}. Each panel is a heat map of maximal efficiency vs the two parameters $\left( \ell_D, \phi_a \right)$, at a fixed tumbling rate $\alpha$. This rate is then changed from panel to panel. 
The first striking result is the strong dependence of efficiency on tumbling rate $\alpha$, with slow tumbling rates leading to more efficient engines (note that the scales for the colour bars in each panel differ in their order of magnitude).  This effect mirrors the behaviour of the ratchet current for $F_l =0$ (considered previously in~\cite{derivaux_rectification_2022-1}) but the effect for the active engines is much stronger. 

\begin{figure}
\includegraphics[width = \textwidth]{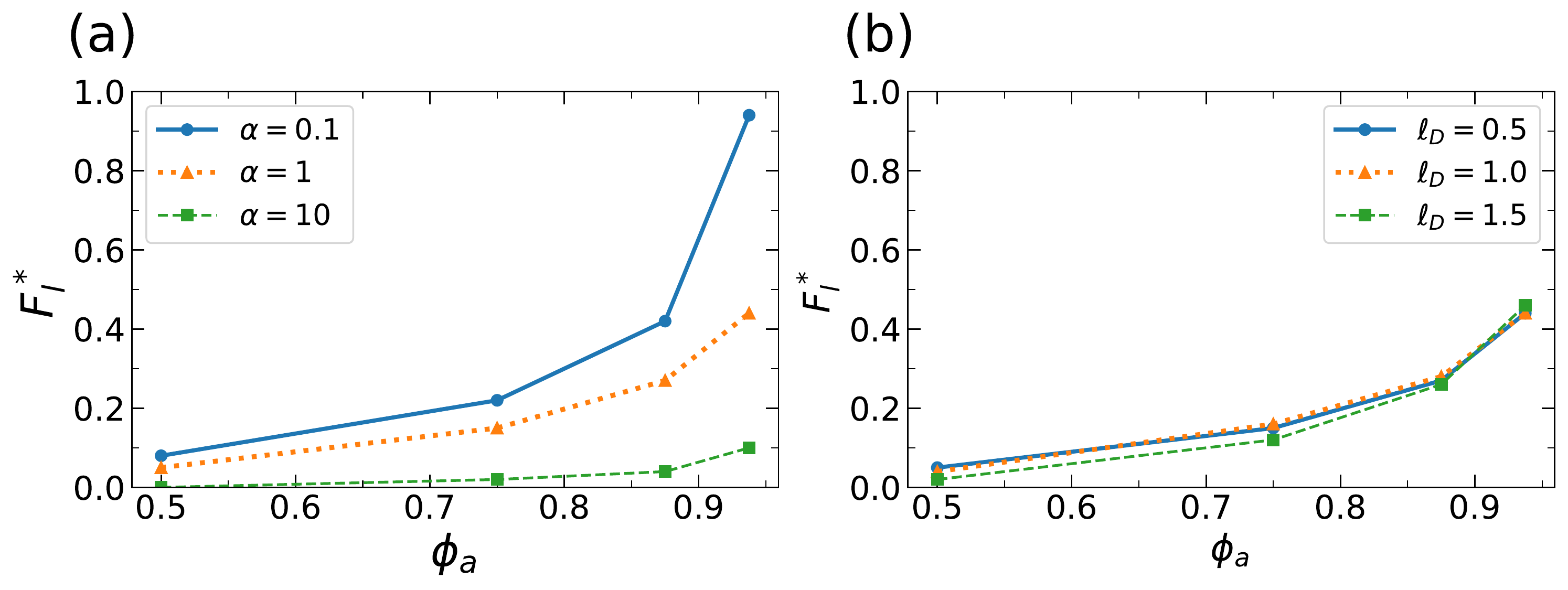}
\caption{Plots of the load force at maximal efficiency. Tumbling $\alpha$ is varied in (a) (with $\ell_D = 0.5$) and the passive diffusivity $\ell_D$ in (b) (with $\alpha = 1$).\\
Parameters: baseline \eqref{equ:baseline} with variation of $\alpha$ and $\ell_D$, $N = 16$. }
\label{fig:Fmaxe_alpha_ellD}
\end{figure}

 Turning to  Fig. \ref{fig:panel_alpha_elld}(a,b), for which $\alpha$ is not too large and the efficiency not too small, we see similar trends in both cases: efficiency tends to be large when $\ell_D$ is small (weak thermal noise) and $\phi_a$ is large (the majority of particles are active).  These results may be expected because large $\ell_D$ was previously observed to reduce the rectification effect~\cite{derivaux_rectification_2022-1}, whereas the enhanced efficiency at large $\phi_a$ was already discussed in Sec.~\ref{sec:composition}.  
 
 It is also notable that for large $\phi_a$ (which tends to be the more efficient regime), the dependence of efficiency on $\ell_D$ is much weaker than its dependence on $\alpha$.  To rationalise this effect, it is useful to consider Fig.~\ref{fig:Fmaxe_alpha_ellD}, which shows the load $F_l$ which achieves the maximum efficiency.  Similar to Fig.~\ref{fig:maxe_theovsnum_panel}, this force increases strongly with $\phi_a$, reflecting the fact that the efficient mode of the engine occurs when a large numbers of active particles push against each  passive one, with a large load force.  The two panels of Fig.~\ref{fig:Fmaxe_alpha_ellD} show that for small values of $\alpha$, active particles can push effectively against very large loads, enhancing this collective effect.  By contrast, reducing the passive noise has little effect on the loads that can be pushed effectively.

\subsection{Summary}
\label{sec:opti-discuss}

We now briefly summarise and discuss the dependence of engine efficiency on model parameters, and their consequences for engine design.

A central observation is that high efficiency is achieved when a large majority of particles are active, and they push in ``teams'' against passive particles with large loads.   An extreme example of this occurs when all particles push against a single macroscopic obstacle, but this is not optimal -- it is preferable instead to have smaller teams, each pushing their own load.  Presumably the excluded volume interactions among particles, which limits how many can fit in a single well of the ratchet potential, reduce the extent to which large numbers of particles can push effectively on a single load.
The effects of geometrical parameters support this idea in that teams which are split across different minima of the sawtooth potential are less effective in pushing.   

In terms of other parameters, it is also desirable to use low overall particle concentrations, although this effect saturates quite quickly. 
Turning to parameters that determines the particle dynamics, we found that small $\alpha$ and small $\ell_D$ both promotes engine efficiency, similar to the situation for the ratchet current in the absence of a load~\cite{derivaux_rectification_2022-1}.
This may be expected: the engine is driven by the active particle motion (which is most pronounced for small $\alpha$), while the thermal fluctuations of the passive particles only hinder efficient engine operation.  Small $\ell_D$ means that these fluctuations are suppressed.

\section{Effects of particle sequence}
\label{sec:sequences}

\subsection{Comparing specific sequences with averaged behaviour}

Recall that since particles cannot pass each other, the sequence of active and passive particles in these systems is conserved under the dynamics, and results such as efficiency depend on the specific sequence.  So far we have shown results obtained by averaging over randomly chosen initial sequences.  We now consider in more detail the effects of the sequence, focussing on the physical consequences for the engine.

\begin{figure}
\includegraphics[width = 0.95 \textwidth]{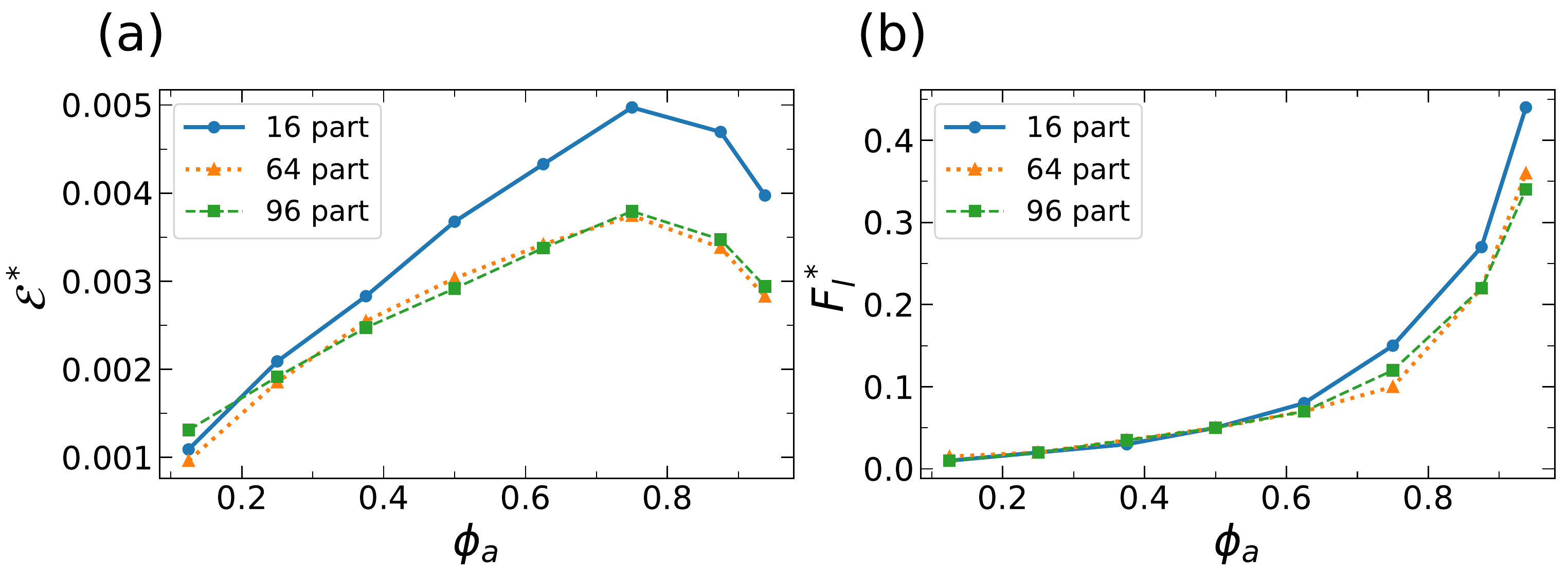}
\caption{(a) The maximum efficiency of the engine, on varying $\phi_a$ and system size $N$. (b) The external force $F_l$ at maximum efficiency. \\
Parameters: as in Fig. \ref{fig:current_activework_scaling_intensive_64part}. }
\label{fig:maxeandF_phia_Npart}
\end{figure}

\begin{figure}
\includegraphics[width = 1.00 \textwidth]{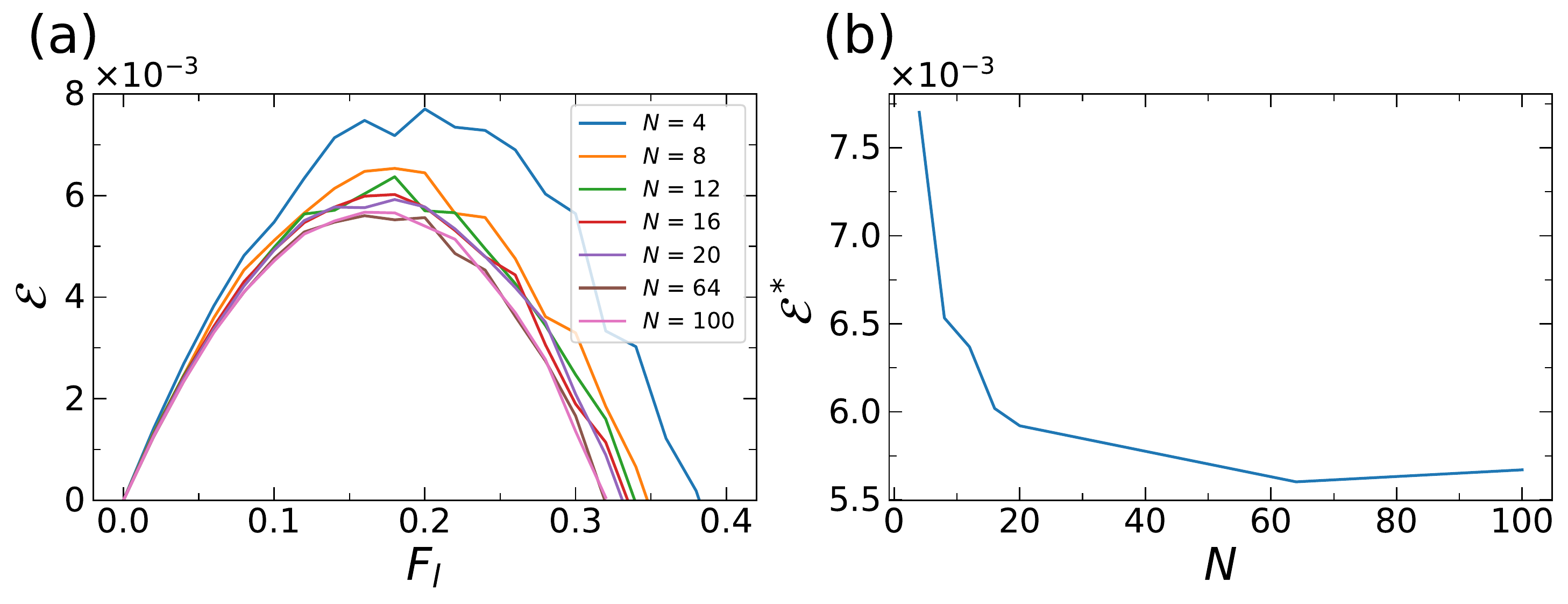}
\caption{ (a) Dependence of efficiency on load force $F_l$ in a system with $\phi_a = 0.75$. The initial sequence for every curve is a $N/4$ repetition of the single unit $0001$. (b) Maximal efficiency for each curve of the left panel plotted against the number of particles $N$. \\
Parameters: baseline Eq.~\eqref{equ:baseline} and  $\phi_a = 0.75$. }
\label{fig:optimalseq_maxe}
\end{figure}

First, Fig.~\ref{fig:maxeandF_phia_Npart} shows the effect of increasing system size $N$, while still averaging over random sequences.  One sees that small systems tend to have higher efficiency, but larger systems seem to converge to results that are independent of $N$.  The increased efficiency in small systems can be partially explained by the example trajectory in Fig.~\ref{fig:panel_tseries_conc}: this starts with a few teams of active particles pushing on passive ones, but these teams eventually accumulate into a single cluster, which seems to moves at the speed of the slowest team.  
The larger is the system, the more likely is the presence of a slow team that hinders the motion of all the others, via the single-file constraint.  Hence smaller systems tend to be more efficient.  In this context, the observation (Fig.~\ref{fig:maxeandF_phia_Npart}) of a limiting efficiency in large systems is important for the scalability of this system. (One might imagine that increasingly large systems might be controlled by increasingly slow teams, leading to ever poorer efficiency; but we find no evidence of this and instead observe saturation.)
 
One way to suppress the effect of rogue slow teams is to use a periodic sequence of particles.  
In the following, we represent the sequences of active/passive particles with a binary code, `0' standing for active and `1' for passive particles.  For $\phi_a=0.75$ (which is efficient for the engine), a suitable sequence is obtained by repeating the four-particle configuration `0001'.  Since each passive particle is then pushed by an identical team of three active ones, there is no opportunity for an anomalously slow team to hinder efficiency.  Results for this sequence are shown in Fig.~\ref{fig:optimalseq_maxe}, for different system sizes.  Comparing with Fig.~\ref{fig:maxeandF_phia_Npart}, the overall efficiency is significantly higher (approximately $0.0055$ compared with $0.0033$), consistent with the hypothesis that efficiency is highest if all teams tend to move equally fast.   The effect of system size on efficiency is also weaker than in  Fig.~\ref{fig:maxeandF_phia_Npart} (note that the comparison is only relevant for $N\geq 16$), consistent with the idea that for random sequences the larger systems in Fig.~\ref{fig:maxeandF_phia_Npart} are being slowed down by the restriction to move at the pace of the slowest team.

To further characterise the slowing effect induced by the initial arrangement, we computed current, active work and efficiency for every possible arrangement in a series of $N = 12$ particles, taking $\phi_a=0.75$ with a fixed load $F_l=0.15$ (these conditions are close to the maximal efficiency obtained from the averaged data of Fig.~\ref{fig:theo_fit_4panels_EB}). Particle size was set to $d = 1/12$ to avoid wells clogging and isolate the effect of arrangement only. The results are tabulated in Appendix \ref{app:tables}.  These data reveal a number of interesting effects, of which we give a brief summary.

\begin{figure}
\includegraphics[width = 1.00 \textwidth]{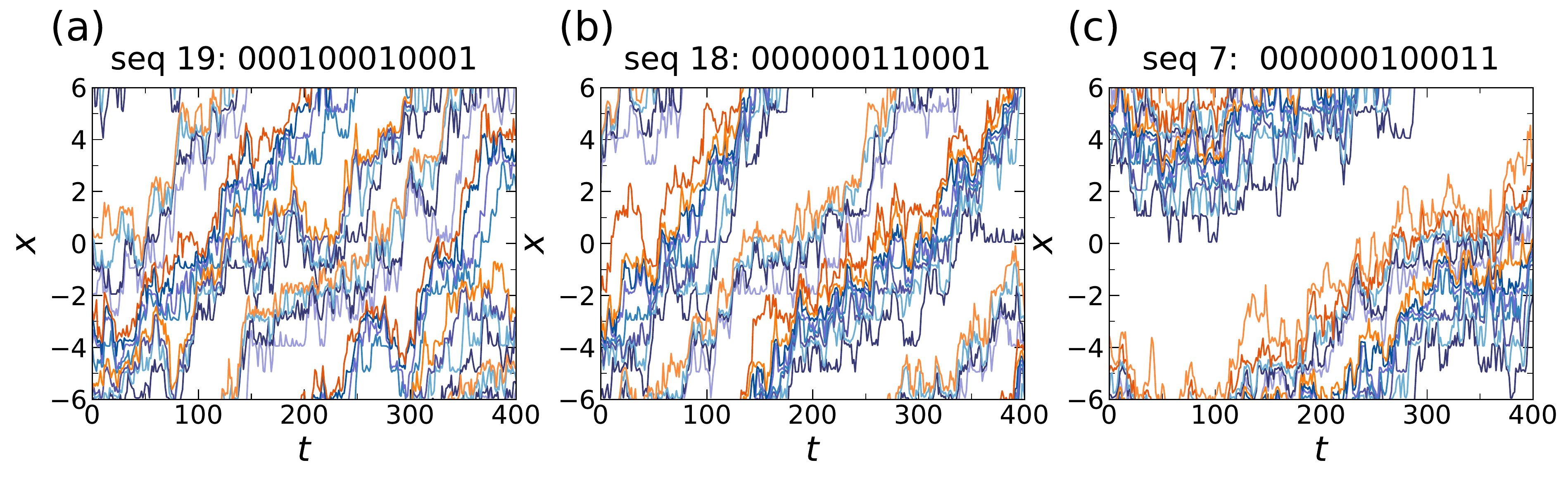}
\caption{Trajectories of various arrangements of an engine of 12 particles with $\phi_a = 0.75$.
for various sequences of active and passive particles, as indicated. (a) Periodic sequence. (b) ``Balanced'' sequence where the larger team of active particles pushes the larger cluster of passive ones.  (c) ``Unbalanced sequence'' where the roles of the larger and smaller clusters are reversed.  The most efficient sequence is that of (a); also (b) is significantly more efficient than (c), see the main text for a detailed discussion.
Parameters: baseline \eqref{equ:baseline} except $\tau = 400$, $t_{\text{eq}} = 1500$ with $\phi_a = 0.75$, $N = 12$, $F_l = 0.15$. }
\label{fig:tseries_3panels_clusters}
\end{figure}

As seen in the Appendix, the efficiencies range from $0.0085$ to $0.01$. Fig.~\ref{fig:tseries_3panels_clusters} shows the behaviour for three illustrative sequences. The sequence with highest efficiency (and highest ratchet velocity) is the periodically-replicated `0001' case considered above.  We compare this with two interesting sequences with intermediate efficiency.  One of them may be abbreviated as $0^61^20^31$: it results in a team 6 active particles pushing a block of two passive ones, together with a team of 3 active particles pushing a single passive one.  In this case the larger active team pushes the larger passive obstacle, and vice versa.  We compare this with $0^610^31^2$ where the situation is reversed: the smaller active team now pushes the larger obstacle.  In the former case, the teams remain separated and move at similar speeds, this sequence is almost as efficient as the optimal one. On the latter case, the smaller team is slower and all particles aggregate into a single cluster that moves at this slower speed, leading to a significantly lower efficiency.  (Note that these two sequences correspond to reflections of each other, once the periodic boundaries are taken into account: the active teams always organise themselves to the left of their corresponding obstacles, so reversing the sequence changes significantly the behaviour.)

The general principles arising from these results are, firstly, that the ratchet velocity tends to be controlled by the slowest team; secondly, that larger teams can still maintain large currents when they push on larger obstacles that contain multiple particles. Sequences of separated teams tend to have a larger active work, which is compensated by a stronger current and leads to overall better efficiency.

\subsection{Effects of particle swapping}
\label{sec:swapping}

As an alternative way to control the particle sequence, we can remove the single-file constraint and allow particles to pass each other within the model.  We implement this by introducing swaps of particle positions when they are in contact as detailed below.  Of course, for the passive particles to move against the load, they must feel significant forces from active particles pushing on them, so this swapping should not happen too frequently, or the engine will not operate effectively.  However, we will show that occasional swaps maintain the operation of the engine, while at the same time allowing the sequence to fluctuate.  If we then perform time averages over long trajectories, then the system will explore many different sequences: this effectively replaces our previous quenched average over sequences by an annealed one, where the systems' dynamical rules determines the relative weights with which each sequence appears.

\begin{figure}
\includegraphics[width = 1.00 \textwidth]{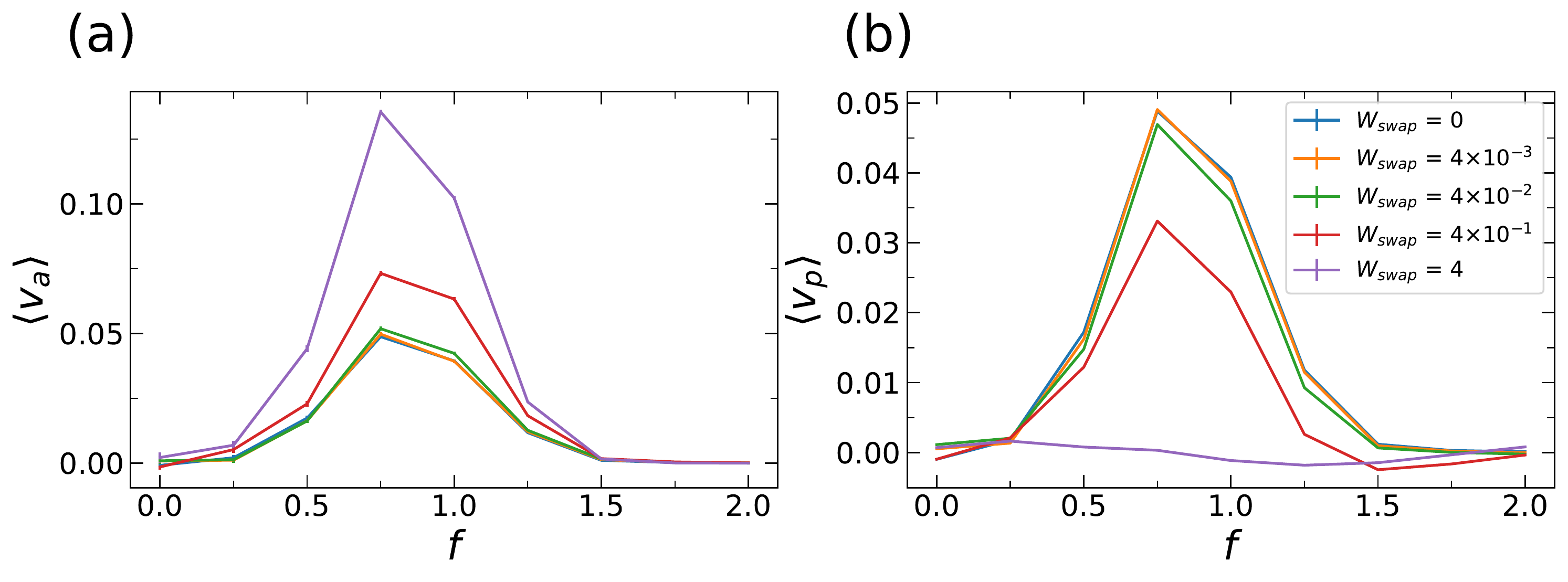}
\caption{Impact of particles swapping on particles velocities. (a) Effect on velocity of active particles. (b)  Effect on velocity of passive particles. \\
Parameters: baseline Eq.~\eqref{equ:baseline} and $\phi_a = 0.5$, $N = 16$, $F_l = 0$. }
\label{fig:swap_APcurrents_f}
\end{figure}

\begin{figure}
\includegraphics[width = 1.00 \textwidth]{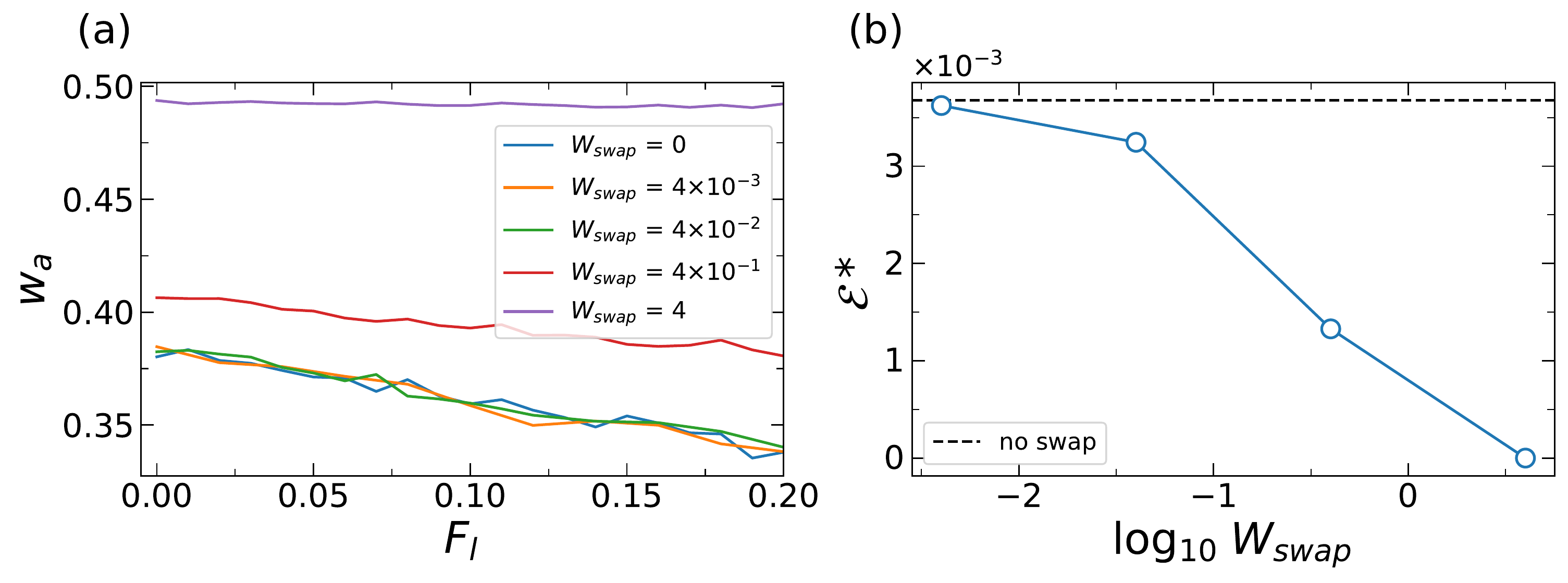}
\caption{(a) Evolution of active work for different values of swap probability rate. (b) Maximal efficiency decreases with larger probability of swapping.  The dashed line shows the result for $W_{\rm swap}=0$.  
\\
Parameters: baseline Eq.~\eqref{equ:baseline} and $\phi_a = 0.5$, $N = 16$. }
\label{fig:swap_panel_wa_maxe}
\end{figure}

To remove the single-file constraint, we allow
 particles to swap by introducing a (stochastic) swap rate $W_{\rm swap}$: if two particles are within a distance $d_{\rm swap}$ of each other, the probability for them to swap in a single time step of duration $\delta t$ is $W_{\rm swap}\delta t$.\footnote{The maximal possible swap rate is then $1/\delta t$: all swap rates considered here are much smaller than this.}  
 (Note these swap rates are independent of the forces acting on the particles.)
 For consistency, we take $d_{\rm swap} = 2^{1/6} d$, which is the same as the cut-off radius of the WCA potential.  
 
When swapping is allowed, passive and active averaged currents, $\langle v_a \rangle$ and $\langle v_p \rangle $, are no longer constrained to be equal.  In general, we find that $\langle v_a \rangle$ increases at high swapping rate due to the less impeded movement of the active particles, while $\langle v_p \rangle$ is expected to decrease due to the lack of rectification.  This trend is confirmed by numerical simulations of a mixture of 16 particles with $\phi_a=0.5$ and without external load, see Fig. \ref{fig:swap_APcurrents_f}. 
That Figure also shows that small swap rates leave currents unchanged, while rates of $W_{swap} \approx 10^{-1}$ start to have an impact, and the passive current becomes negligible for $W_{swap} \approx 1$.

Having verified that swapping operates as expected in the absence of any load, we return to the engine operation with $F_l>0$.  Given the previous results, we expect that swapping will reduce the passive particle velocity and hence the efficiency, but this effect should not be too strong for small swap rates.  One also expects an increase in the active work, because the motion of the active particles is less hindered by the passive ones.  These expectations are confirmed by Fig.~\ref{fig:swap_panel_wa_maxe}.

\section{Conclusion}
\label{sec:conclusions}

We have proposed an active engine in one dimension where the left-right symmetry is broken by the interaction between active particles and an external potential, and work is extracted from isotropic passive particles with which the active ones collide.  This may be contrasted with approaches where active particles do work on asymmetric obstacles~\cite{pietzonka_autonomous_2019,sokolov_swimming_2010,leonardo_bacterial_2010,vizsnyiczai_light_2017}.  By separating the anisotropy from the work extraction, our approach lifts some of the constraints inherent in previous designs, which typically assume (for example) that the passive components (obstacles) are much larger than the active ones.  The model also has connections to previous work on molecular motors~\cite{julicher_modeling_1997}.

The performance of the active engine was assessed through the engine efficiency $\mathcal{E}$.  Numerical simulations showed that composition of the mixture is a central quantity for optimisation of efficiency. A linear approximation for the current $\langle v \rangle$ qualitatively reproduces  the variation of efficiency due to composition change.  The best efficiency is found at a relatively large $\phi_a$, where active particles can self-organise into a team, to push a single passive one.  However, very large $\phi_a$ leads to less efficient work extraction, when there are too few passive particles to push efficiently. 

In addition to the mixture composition, we investigated the impact of particles' dynamics and their geometrical parameters on efficiency.
Overall, the parameters that achieve high efficiency are those where rectification without load is also efficient~\cite{derivaux_rectification_2022-1}, although an important consideration in the engine operation is that the loads on passive particles mean that they strongly impede the motion of active ones, which favours self-organisation into clusters.
Indeed, we find that maximal efficiency is obtained when small teams of active particles act cooperatively to push a single passive particle. By investigation of specific sequences of active and passive particles (which are preserved under the dynamics), we showed that this mechanism is particularly effective if teams can evolve independently without impeding each other: this occurs, for example, in some periodic sequences.

Given its simple one-dimensional structure, the model is mostly useful for illustrating general principles for engine design. As a first such principle, note that the engine operation relies on a ratchet effect, which can be expected in systems where spatial inversion symmetry is broken by the ratchet potential, and time-reversal symmetry is broken by self-propulsive forces.  Hence, one may expect the qualitative behaviour observed here to be robust, even if the active particles had different dynamics such as active Ornstein-Uhlenbeck particles \cite{martin_statistical_2021} or active Brownian particles \cite{fily_athermal_2012,redner_structure_2013-1}, or other active models \cite{schweitzer_complex_1998}.  It would be interesting to explore how engine efficiencies vary between different kinds of active particle.  (For example, the fact that run-and-tumble particles feel constant (identical) propulsive forces may be relevant for their self-organisation into teams.)

The single-file motion in this model is a one-dimensional feature that is not generic for active engines, since particles may squeeze past each other in higher dimensional systems, even if the density is high.  However, the results of Sec.~\ref{sec:swapping} show that breaking the single-file constraint does preserve the main features of the engine, which helps to justify our approach. A two-dimensional extension of the sawtooth potential has been also shown to rectify active particles \cite{mcdermott_collective_2016} despite their ability to bypass each other. Those both elements indicate that rectification of passive particles and active work extraction seems generalisable to higher dimensions. An other interesting feature of higher dimensions is that anisotropy of active particles can also promote alignment \cite{solon_pressure_2015-1} which is absent in the current model and might have interesting consequences for engine efficiency.

Overall, our model proves that such engines can operate effectively, and that self-organised clusters/teams of active particles can aid efficiency. We have not attempted a global optimisation of all parameters, but favourable conditions (for example, $\phi_a = 14/16$, $ d=0.05$, $c = 0.5$, $\alpha = 0.1$, $\ell_D =0.5$) result in a maximal efficiency $\approx 5\%$. This value is close to the one obtained with asymmetric obstacles ($\approx 8\%$) \cite{pietzonka_autonomous_2019}, indicating that these different active engines can achieve comparable efficiencies, despite their different design principles.

Taking advantage of the simplicity of this (one-dimensional) model, it would be interesting to attempt a deeper theoretical analysis, to understand particular limits and scaling regimes (for example, in the thermodynamic limit of large system size, or cases where $\phi_a$ is very large so that there are very few passive particles). Theoretical work could be extended further by deriving analytical expression of $\langle v \rangle$ and $w_a$ for a pair of active and passive particles. This derivation would eventually lead to direct analytical expressions relating efficiency to important model parameters such as tumbling rate and particles size. Assessing how these results would be generic to alternative engine designs would prove valuable. As noted above, generalisation to higher dimensions would also be possible.  More practically, experimental realisations of such engines would seem to be possible, for example in microfluidic devices \cite{caballero_motion_2016,katuri_directed_2018,williams_confinement-induced_2022-1,locatelli_single-file_2016,chupeau_optimizing_2020}.
Ultimately, we hope these studies will contribute to the conceptual toolbox of controlled transport and work extraction in mixtures of active and passive objects.

\subsection*{Acknowledgments}
We thank Tirthankar Banerjee, Camille Scalliet and Liheng Yao for useful discussions. This work was funded in part by the European Research Council (ERC) under the EU's Horizon 2020 Programme, Grant agreement No. 740269.
MEC was funded by the Royal Society.

\appendix
\section{Results for individual sequences of active and passive particles}
\label{app:tables}

\begin{table}[h]
\caption{\label{tab:sequences_phia0.75} Ratchet current, active work and efficiencies for the 19 unique sequences in a system of $N_a = 9, N_p = 3$ ($\phi_a = 0.75$).\\
Parameters: baseline Eq.~\eqref{equ:baseline} except $d=1/12$, $\tau = 1200$, $t_{\rm eq} = 300$ and $F_l=0.15$. }

\begin{ruledtabular}
\begin{tabular}{ccccc}
Sequence  &Sequence & $\langle v \rangle $ & $w_a$ & $\mathcal{E} $ \\

\# &  & $(\times 10^{-2})$ & & $(\times 10^{-2})$ \\
\hline
  1 & 000000010101 & 4.08 & 0.241 & 0.845 \\
  2 & 000000001011 & 3.69  & 0.214 & 0.864 \\
  3 & 000000010011 & 3.92 & 0.222 & 0.883 \\
  4 & 000000100101 & 4.46 & 0.249 & 0.895 \\
  5 & 000001000101 & 4.58 & 0.254 & 0.902 \\
  6 & 000000001101 & 4.15 & 0.225 & 0.922 \\
  7 & 000000100011 & 4.18 & 0.226 & 0.925 \\
  8 & 000000101001 & 4.82  & 0.259 & 0.931 \\
  9 & 000010000101 & 4.84 & 0.257 & 0.941 \\
 10 & 000000000111 & 3.80 & 0.200   & 0.947 \\
 11 & 000001001001 & 5.15 & 0.267 & 0.964 \\
 12 & 000001000011 & 4.45 & 0.230  & 0.969 \\
 13 & 000001010001 & 5.09 & 0.261 & 0.975 \\
 14 & 000001100001 & 4.63 & 0.235 & 0.986 \\
 15 & 000000011001 & 4.76 & 0.241 & 0.990  \\
 16 & 000010010001 & 5.59 & 0.273 & 1.024 \\
 17 & 000010001001 & 5.60 & 0.273 & 1.025 \\
 18 & 000000110001 & 4.93 & 0.240  & 1.026 \\
 19 & 000100010001 & 5.80 & 0.277 & 1.046 \\
\end{tabular}
\end{ruledtabular}

\end{table}

As discussed in the main text, the single-file constraint in this system means that the ordering of active and passive particles is preserved by the dynamics.  For the majority of our results, we averaged over several random sequences.  To investigate how results vary between sequences, Table~\ref{tab:sequences_phia0.75} shows results for all possible initial sequences in a system with 9 active particles and 3 passive ones.  See also Fig.~\ref{fig:tseries_3panels_clusters} and the associated discussion.

\section{Dependence of active work on $\phi_a$}
\label{app:activework}

In Section \ref{sec:composition}, we made the assumption that active work depends weakly on $\phi_a$ (compared to its dependence on $\langle v \rangle$): 
\begin{equation}
w_a \approx \langle w_a \rangle_0 \approx \text{const}.
\end{equation} 
This assumption simplifies the optimisation of efficiency with respect to $\left( \phi_a, F_l \right)$. In this appendix, we briefly assess the impact of relaxing this assumption on the approximate theoretical results of Sec.~\ref{sec:composition}.

The theoretical efficiency $\mathcal{E}^{\mathrm{th}}$ is obtained by plugging the formula \eqref{eq:v-model} for the ratchet current and assuming that $w_a$ is constant, which results in the dashed lines of Fig. \ref{fig:theo_fit_4panels_EB}, \textit{ie}
\begin{equation}
\mathcal{E}^{\mathrm{th}} = F_l \times \left[\frac{r\, \phi_a - p\, F_l - q\, (1-\phi_a) \,F_l }{ w_a  } \right] \times \frac{1-\phi_a}{\phi_a} \,.
\label{eq:theo_eff}
\end{equation} 
where $w_a$ is taken to be its value in the purely active system, $\phi_a=1$.

\begin{figure}
\includegraphics[width = 1.00 \textwidth]{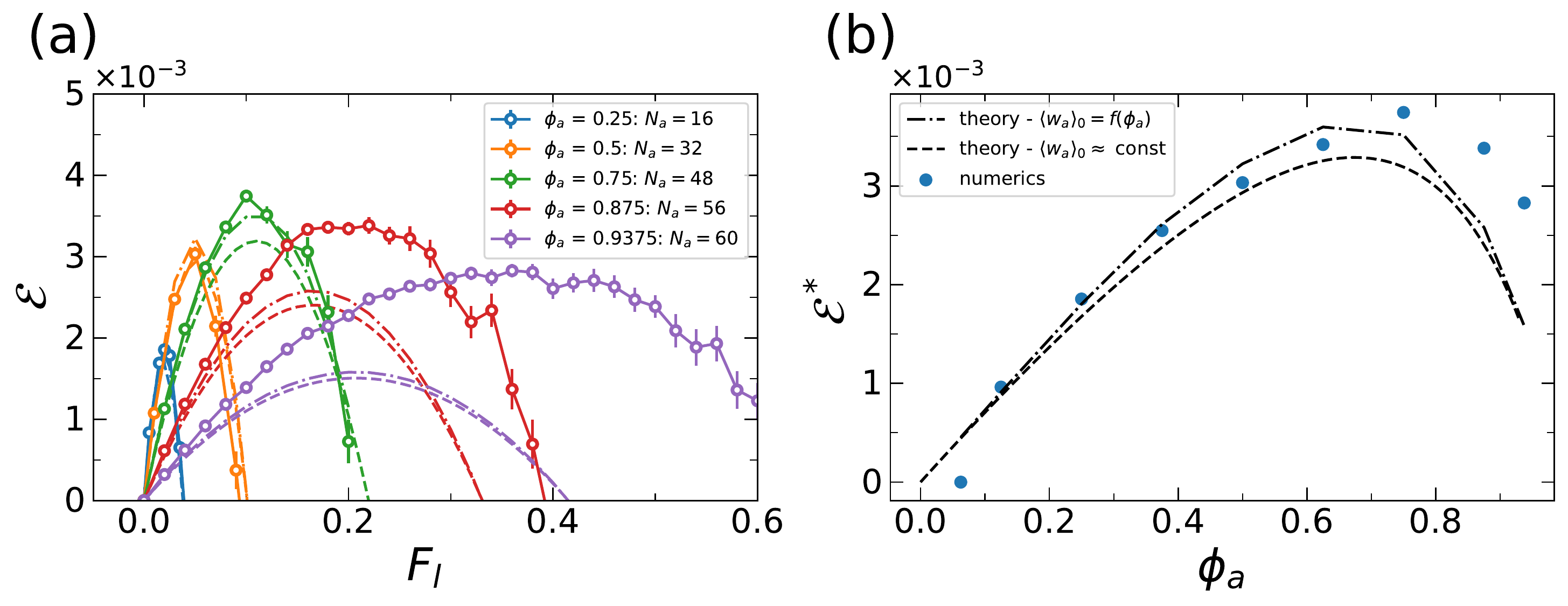}
\caption{(a) Engine efficiency as a function of $\phi_a$.  
Dots represent data from numerical simulations, dashed curves show the simple theoretical approximation~\eqref{eq:theo_eff} , and dashed-dotted curves show the improved approximation~\eqref{eq:theo_mod}.  This improvement from this modified theory is weak.
(b) The maximal efficiency as a function of $\phi_a$.
\\
Parameters: Parameters: baseline Eq.~\eqref{equ:baseline} and $N = 64$. }
\label{fig:app_activework}
\end{figure}

To assess the impact of the  $\phi_a$-dependence of $w_a$, we computed a modified theoretical efficiency  by keeping the same model for the ratchet current but extracting the value of $w_a$ at each $\phi_a$.
Results are shown in Fig.~\ref{fig:app_activework}. Dashed lines represent the theoretical efficiency  \eqref{eq:theo_eff} while the dotted-dashed curves represent the theoretical efficiency with active work now depending on $\phi_a$, $w_a = f(\phi_a)$ and computed from the simulations:
\begin{equation}
\mathcal{E}^{\prime} = F_l \times \left[ \frac{r\, \phi_a - p\, F_l - q\, (1-\phi_a) \,F_l }{f(\phi_a)} \right] \times \frac{1-\phi_a}{\phi_a} \,.
\label{eq:theo_mod}
\end{equation} 
 Fig.~\ref{fig:app_activework}(a) shows that accounting for the $\phi_a$-dependence of $w_a$ only
slightly improves the accuracy of efficiency curves. The effect on the maximal efficiency is also weak, see Fig.~\ref{fig:app_activework}(b).  

\bibliographystyle{apsrev4-2}
\bibliography{Offline}

\end{document}